\pdfoutput=1

\RequirePackage{lineno}

\documentclass{tcibook}
\usepackage{fancyhea}
\usepackage{work}
\usepackage{bm}       
\usepackage{graphics}
\usepackage{amsmath}
\usepackage{graphicx}
\usepackage{hyperref}      
\usepackage{subfigure}

\newcommand{\nc}{\newcommand}  

\def\beq{\begin{equation}}
\def\eeq#1{\label{#1}\end{equation}}
\def\eeqn{\end{equation}}

\newenvironment{Eqnarray}%
   {\arraycolsep 0.14em\begin{eqnarray}}{\end{eqnarray}}
\def\beqa{\begin{Eqnarray}}
\def\eeqa#1{\label{#1}\end{Eqnarray}}
\def\eeqan{\end{Eqnarray}}

\nc{\ra}{\rightarrow}  
\nc{\slsh}{\slash\hspace*{-0.22cm}}
\def\Re{{\cal R \mskip-4mu \lower.1ex \hbox{\it e}\,}}
\def\Im{{\cal I \mskip-5mu \lower.1ex \hbox{\it m}\,}}

\nc{\vev}[1]{ \left\langle {#1} \right\rangle }
\nc{\bra}[1]{ \langle {#1} | }
\nc{\ket}[1]{ | {#1} \rangle }
\nc{\fb}{\,{\rm fb}^{-1}}
\nc{\ev}{{\rm eV}}
\nc{\kev}{{\rm keV}}
\nc{\Mev}{{\rm MeV}}
\nc{\gev}{{\rm GeV}}
\nc{\tev}{{\rm TeV}}
\nc{\mev}{{\rm MeV}}

\def\del{\partial}
\def\Dslash{\not{\hbox{\kern-4pt $D$}}}
\def\dslash{\not{\hbox{\kern-2pt $\del$}}}
\def\pslash{\not{\hbox{\kern-2pt $p$}}}
\def\ETmiss{ \not{\hbox{\kern-4pt $E$}}_T }

\def\msb{{\bar{\ssstyle M \kern -1pt S}}}

\begin{document}

\def\bibname{References}
\bibliographystyle{hunsrt}

\raggedbottom

\pagenumbering{roman}

\parindent=0pt
\parskip=8pt
\setlength{\evensidemargin}{0pt}
\setlength{\oddsidemargin}{0pt}
\setlength{\marginparsep}{0.0in}
\setlength{\marginparwidth}{0.0in}
\marginparpush=0pt


\pagenumbering{arabic}

\renewcommand{\chapname}{chap:intro_}
\renewcommand{\chapterdir}{.}
\renewcommand{\arraystretch}{1.25}
\addtolength{\arraycolsep}{-3pt}



 
\chapter{Top quark working group report}
\label{chap:top}

\begin{center}\begin{boldmath}



\begin{center}

\begin{large} {\bf Conveners: K.~Agashe, R.~Erbacher, C.~E.~Gerber, K.~Melnikov, R.~Schwienhorst.} \end{large}

{\bf Contacts by topic:}
A.~Mitov, M.~Vos, S.~Wimpenny (top mass);
J.~Adelman, M.~Baumgart
% ...
J.~Zupan.

\end{center}



\end{boldmath}\end{center}


\section{Introduction} 
\label{sec:topintro}

The top quark was discovered in 1995~\cite{Abe:1995hr,Abachi:1995iq} and it is still the heaviest
elementary particle known today. Thanks to its large mass, and the related strength of its
coupling to the Higgs boson, the top quark may be a key player in
understanding the details of electroweak symmetry breaking.
Studies of the top quark properties at the Tevatron and Run I of the LHC have given
us a detailed understanding of many properties of this particle, including its mass,
production and decay mechanisms, electric charge and more. 
With the exception of the large forward-backward asymmetry in $t \bar t$ production that
has been observed at the Tevatron, all results on top quark pairs and single top production
obtained so far have been consistent with the Standard Model. We note that in this context, 
the anomaly in the $b$ quark forward-backward asymmetry observed at LEP might get amplified for the much heavier top quark.

In the short and mid-term future, top quark studies will be mainly driven by the LHC experiments. 
Exploration of top quarks will, however, be an integral part of particle physics studies at any
future facility.  Future lepton colliders will have a rich top quark physics program which would
add to our understanding of this interesting quark.
Detailed simulation studies have been carried out for linear electron-positron machines
(ILC and CLIC). First attempts have been made to extrapolate these to the case of a circular
machine (TLEP). In this report we describe what can be achieved based on projection studies 
for the LHC and for future lepton colliders. The report is organized along six topics:

\begin{itemize} 
\item Measurement of the top quark mass;
\item Studies of kinematic distributions of top-like final states; 
\item Measurements of top quark couplings; 
\item Searches for rare decays of top quarks;
\item Probing physics beyond the Standard Model with top quarks; 
\item Algorithms and detectors for top quark identification at future facilities. 
\end{itemize}

Main conclusions for each topic are presented in Sect.\ref{sec:topconclusions}.

\section{The top quark mass} 
\label{sec:topmass}
~
The top quark mass is a parameter whose precise value 
is essential for testing the overall consistency of the Standard Model or 
models of new physics through precision electroweak fits. 
The exact value of the top quark mass is also crucial for understanding whether the Standard Model
{\it without further extensions}  can be continued to energies compared to the Planck scale,  
without running into problems with the stability of electroweak  vacuum \cite{Degrassi:2012ry}.  
To put both of these statements into perspective, we note that 
the value of the top quark mass, as quoted by the Particle Data Group, is $m_t = 173.5 \pm 0.6 \pm 0.8~{\rm GeV}$. 
The total uncertainty on $m_t$ is therefore close to $1~{\rm GeV}$; this is the best relative 
precision available for {\it any} of the  quark masses. 

Nevertheless, we know that for precision electroweak fits, a $0.9~{\rm GeV}$ uncertainty 
in the top quark mass corresponds to a $5.4~{\rm MeV}$ uncertainty in the $W$-mass
(see e.g. Refs.~\cite{pew} and~\cite{Baak:2012kk}).  Since  the $W$-mass is expected to be  
measured with this precision at the LHC, and significant improvements in $\delta M_W$ 
beyond this are not likely, we conclude that the future of
precision electroweak physics requires the measurement of the top quark mass
to at least a precision of less than $0.5~{\rm GeV}$, and desireably to 0.3~GeV so that the
top sector is not limiting in EW precision fits.

In addition, the vacuum stability issue depends 
strongly on the value of the top quark mass. Indeed, as shown in Ref.~\cite{Degrassi:2012ry},  
changing $m_t$ by $2.1~{\rm GeV}$ around the central value $m_t = 173.1~{\rm GeV}$, 
the energy scale where the Higgs potential becomes unstable changes 
by {\it six} orders of magnitude, from $\mu_{\rm neg} \sim 10^8~{\rm GeV}$ 
to $\mu_{\rm neg} \sim 10^{14}~{\rm GeV}$!  
It is easy to estimate that if $m_t$ is known 
with $0.3-0.5~{\rm GeV}$ uncertainty, as required by  the electroweak fit, the scale can 
be estimated much more precisely, to within a factor of five. 
We conclude that the knowledge of the top quark mass with the $0.5~{\rm GeV}$  uncertainty will have an 
important impact on our understanding of particle physics. 

Furthermore, it has recently been suggested~\cite{tlep} that a much more precise 
measurement of the $W$ mass can be performed at a  
circular $e^+e^-$ collider such as TLEP, where $\delta M_W \le 1.5~{\rm MeV}$ can probably be achieved. 
For the purpose of precision electroweak 
fits, such high precision can be only utilized if the top quark mass is measured with the matching precision of about $0.1~{\rm GeV}$.
As we explain below, this can be  accomplished at an $e^+e^-$ collider such as the ILC, CLIC, 
or TLEP itself. Knowing
$m_t$ with such a precision will also allow for a much more decisive tests  of the vacuum 
stability problem in the Standard Model. 
The interest in testing this scenario 
may increase greatly if no new physics at the TeV scale is found in the Run II of the LHC. 
Note that for these purposes, a numerical value for  
theoretically well-defined top quark mass parameter, for example $m_t^{\overline {\rm MS}}$,  is required.

\subsection{Linear Colliders}

A $e^+e^-$  collider will allow us to study  electroweak production of $t\overline{t}$ 
pairs with no concurring QCD background.
Therefore, precise measurements of top quark properties become possible. 

\begin{figure}
\centering
  \includegraphics[width=0.49\textwidth]{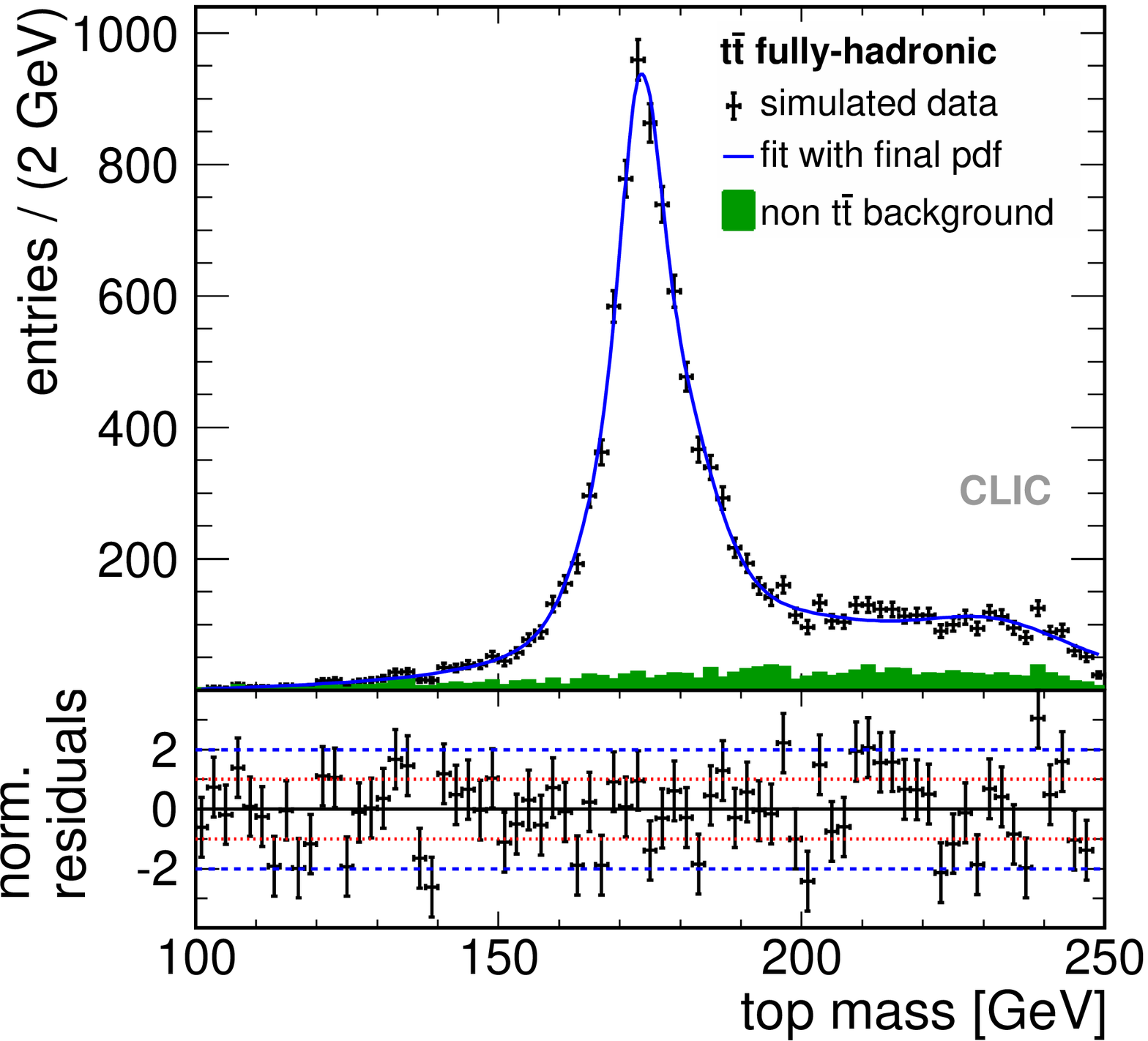}
   \includegraphics[width=0.49\textwidth]{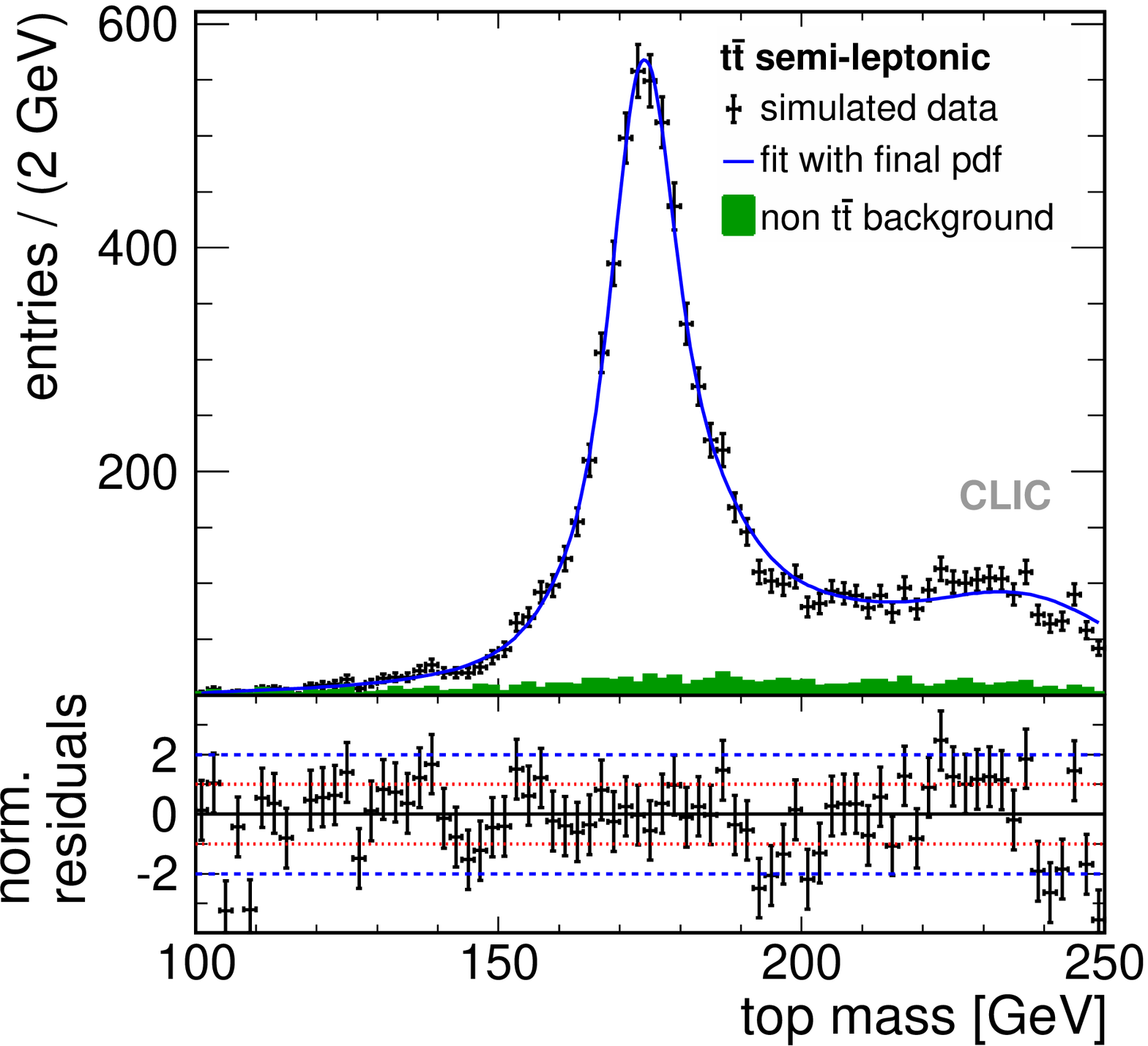}
        \caption{Distribution of reconstructed top mass for events classified as fully-hadronic (left) and semileptonic (right). The data points include signal and background for an integrated luminosity of 100\,fb$^{-1}$~\cite{Seidel:2013sqa}. The pure background contribution contained in the global distribution is shown by the green solid histogram. The top mass is determined with an unbinned likelihood fit of this distribution, which is shown by the solid line.}
   \label{fig:TopMass}
 \end{figure}

The top quark mass can be measured at  $e^+e^-$ machines using two complementary methods. 
First, one can use  the invariant mass of the reconstructed $bW$ system from the top decay.
The result of a full simulation study at a 500~GeV linear collider~\cite{Seidel:2013sqa} (CLIC,
with similar results for ILC) is shown in Fig.~\ref{fig:TopMass}. 
The figure demonstrates also the small residual background expected for top quark studies at 
any $e^+e^-$ machines. In the second 
method the top mass is determined in a threshold scan, an option unique to an $e^+e^-$ machine. 
In the threshold scan the so-called 1S top quark mass can be measured to an experimental precision of better than
40~MeV where studies have shown that the statistical error is dominant. Expressing the measurement in terms of 
the theoretically well defined \mbox{$\overline{\rm MS}$} mass will inflate the uncertainty 
to $\sim100$~MeV, as shown in detailed simulations~\cite{Martinez:2002st, Seidel:2013sqa,Asner:2013hla}
and advanced theoretical computations ( see e.g. Ref.~\cite{Beneke:2008ec} and references therein). 

We note that with respect to the top quark mass determination, all lepton colliders that were
suggested so far perform similarly\footnote{We note that some improvements in the $m_t$ 
determination can be expected at the muon collider and at TLEP thanks to reduced beamstrahlung,
although this still has to be demonstrated by detailed simulations.}
and that an additional attraction of measuring $m_t$ at a lepton collider is a clean theoretical 
interpretation of the result of the measurement. As we explain below, the situation is  
more confusing at a hadron collider,  although new methods 
for $m_t$ measurements developed at the LHC help to mitigate this difference.


\subsection{Top quark mass at the LHC}

As previously noted, a precision of $0.5~{\rm GeV}$ or better  in the top quark mass 
is motivated by the future of precision electroweak fits.  It is an interesting question 
whether $m_t$ measurements with such 
a precision  can be accomplished 
at the LHC. To answer it, we will first make some general remarks about measurements of $m_t$. 

Existing  measurements of the top quark mass 
rely  on  complex techniques  required by the difficult hadron collider environment.
The highest accuracy is  currently achieved using the so-called matrix-element 
method ( for a recent review, see   \cite{Gainer:2013iya}).  
We will explain a  generic measurement of the  top quark mass  by considering the following example. 
Any measurement of the top quark mass is based on fitting a particular piece of data to a theory 
prediction where $m_t$ enters as a free parameter. Hence, we write 
\begin{equation}
D = T(m_t,\alpha_s,\Lambda_{\rm QCD} )  = T^{(0)}(m_t) + \frac{\alpha_s}{\pi}  T^{(1)}(m_t) + {\cal O}(\Lambda_{\rm QCD}/m_t,\alpha_s^2),  
\label{eq0}
\end{equation}
where $D$ on the left hand side is a particular kinematic distribution measured in experiment and $T$ on the right-hand side is a theoretical 
prediction, expanded in power series in the strong coupling constant.   
We have indicated in Eq.(\ref{eq0}) that the selected distribution should be minimally affected by 
non-perturbative corrections; we will return to this point below. We also note that inclusion of QCD corrections 
necessitates a clear definition of the renormalization scheme, which then fixes the mass parameter extracted from 
the fit. Since the two popular choices of the renormalized mass parameter, the pole mass and the $\overline {\rm MS}$ mass,  
differ by almost $7~{\rm GeV}$, the specification of the renormalization scheme in the extraction of the top quark 
mass is an important issue.  Solving Eq.(\ref{eq0}), we find the top quark mass $m_t$.  In general, the quality 
of such a solution 
depends on the {\it accuracy} of the theoretical prediction that we have in the right hand side, which is 
controlled by the order in perturbation theory included there.  The majority 
of current  analyses are performed with 
leading order theoretical tools. This amounts to setting $T^{(1)} \to 0$ in the above equation. 
The expected error on $m_t$ is then
\begin{equation}
\delta m_t \sim \frac{\alpha_s}{\pi} \frac{T^{(1)}}{T^{(0)'} } 
\sim \frac{\alpha_s m_t}{\pi}   \frac{T^{(1)}}{T^{(0)}}
\sim \frac{\alpha_s}{\pi}  m_t \sim 6~{\rm GeV}, 
\label{eq1}
\end{equation}
where $T^{(0)'} = {\rm d} T^{(0)}/{\rm d} m_t$ and we used $T^{(0)'} \approx T^{(0)}/m_t$.  It is obvious 
from Eq.(\ref{eq1}) that the estimated error in Eq.(\ref{eq1}) 
is {\it significantly larger} than the current ${\cal O}(1)~{\rm GeV}$ error on $m_t$. 
We conclude that if $m_t$  is obtained from a generic distribution at leading order, one can 
not, in general, expect the accuracy that is better than few GeV.  Fortunately, there are two ways to get around this problem.
The first one requires  inclusion of  NLO QCD corrections into a theory prediction; effectively, this  pushes  the error to 
$m_t (\alpha_s/\pi)^2  \sim 0.3 ~{\rm GeV}$ which is acceptable. The second one  amounts to 
finding  a kinematic distribution  which has a {\it strong} dependence 
on $m_t$; in this case,   ${\rm d}T^{(0)}/{\rm d} m \gg T^{(0)}/m_t$ and the estimate in Eq.(\ref{eq1}) 
receives an additional suppression. 

As we show below, {\it new}  
experimental techniques  that   address the question of the  
top quark mass determination  follow the two approaches described above. 
The above discussion can be used to argue 
that {\it well-established}  methods for the top quark mass determination may have additional 
systematic errors which are not accounted for in their error budgets. 
Indeed, the matrix element method\footnote{The template method \cite{ATLAS:2012aj} is subject to similar arguments.}  
is designed to  maximize  probabilities for  
kinematics of observed events by 
adjusting values of the top quark mass on an event-by-event basis. 
It can be thought therefore as an attempt to fit a very large 
number of kinematic distributions for the best value of $m_t$. 
An unsatisfactory feature of this method is its ``black-box'' nature that 
does not allow one to understand which kinematic features of the top quark pair production process drive this sensitivity.
While such methods -- by design -- should find distributions that show strong  dependence on $m_t$, 
it is not clear if the relevant distributions are  sensitive to non-perturbative effects whose description 
from first principles is not possible.  Moreover, such approaches routinely rely on the use of parton 
shower event generators instead of proper QCD theory. This means that  Eq.(\ref{eq0})  becomes 
\begin{equation}
D = T(m_t,\alpha_s,\Lambda_{\rm QCD}) \approx T^{(0)}_{\rm MC}(m_t,\alpha_s,\Lambda_{\rm QCD},
{\rm tunes}), 
\label{eq2} 
\end{equation}
where, as indicated in the last step, additional approximations, including parton shower tunes, are performed on the ``theory'' side. 
While the quality of this approximation 
{\it for the purpose of determining $m_t$} may be good, 
it is simply not clear how to assign the error 
to the parameter $m_t$ which is extracted following this procedure. 
To make this problem explicit, the top quark mass extracted from Eq.(\ref{eq2}) should be properly  referred to 
as the ``Monte-Carlo 
mass '', whose relation to short-distance definition 
of $m_t$  is not understood.

\begin{table}[t]
\begin{center}
\begin{tabular}{|c|c|c|c|c|c|c|c|c|}
\hline
& {\rm Ref.}\cite{Chatrchyan:2012cz} & \multicolumn{5}{c|} {\rm Projections}\\
\hline
{\rm CM Energy} & {\rm 7 TeV } & \multicolumn{5}{c|}{\rm 14 TeV  } \\
\hline
{\rm Luminosity} & {$5 fb^{-1}$} & \multicolumn{2}{c|}{$100 fb^{-1}$} & \multicolumn{2}{c|}{$300 fb^{-1}$} & {$3000 fb^{-1}$} \\ 
\hline
{\rm Pileup} & 9.3 & 19 & 30 & 19 & 30  & 95  \\
\hline\hline
{\rm Syst. (GeV)} & 0.95 & 0.7 & 0.7 & 0.6 & 0.6 & 0.6 \\
\hline
{\rm Stat. (GeV)} & 0.43 & 0.04 & 0.04 & 0.03 & 0.03 & 0.01 \\
\hline\hline
{\bf Total, GeV } & {\bf 1.04} & {\bf 0.7} & {\bf 0.7} & {\bf 0.6} & {\bf 0.6} & {\bf 0.6} \\
\hline
\end{tabular}
\vspace*{0.5cm}
\caption{Precision of the 
top quark mass measurements that can be expected using conventional (likelihood-type) methods. 
Extrapolations are based on the published CMS lepton-plus-jets analysis. An additional $0.3~{\rm GeV}$ 
systematic error was  added to all extrapolated results.}
\label{Table_massconv}
\end{center}
\end{table}

In spite of the caveats with  the top quark mass  determination that are inherent to  conventional methods, 
it is interesting  to estimate precision in $m_t$ that can be achieved at the LHC. 
We do that using extrapolations of what has been accomplished at the Tevatron and during  the run I of the LHC. 
In Table~\ref{Table_massconv} we show such projections for conventional methods assuming that the 
mass is measured in the lepton + jet channel for the $14~{\rm TeV}$ LHC for different integrated luminosities 
and pile-up scenarios.  We assume the $t \bar t$ production 
cross-section to be $\sigma_{pp \to t \bar t} = 167(951)~{\rm pb}$ at $7$ and $14$ TeV LHC, respectively. 
It follows from Table~\ref{Table_massconv} that conventional 
methods may, eventually,  lead  to the measurement of  the top quark mass with an  error of about $0.6~{\rm GeV}$ and 
that this error is totally dominated by systematic uncertainties. 
It is interesting to point out  that precision in $m_t$ saturates for the integrated luminosity of $300~{\rm fb}^{-1}$ 
and that there is no benefit of using yet higher luminosity  for the top quark mass measurement. 
The reason for this is the increased pile-up and related degradation 
of the jet energy scale determination  in the high-luminosity environment, see a detailed 
discussion in Section~\ref{sec:unboosted}. Note, however, that the systematic 
error estimate in Table~\ref{Table_massconv} includes $0.3~{\rm GeV}$  that was  added to all extrapolated 
results to account for unforeseen sources of systematics; if we omit  this $0.3~{\rm GeV}$ uncertainty, the 
uncertainty on the top quark mass measurement becomes very small.

Conceptual problems with conventional methods  can be mitigated be measuring the top quark mass from 
well-defined kinematic distributions which, on the one hand, 
are sufficiently sensitive to $m_t$ and, on the other hand, can be cleanly interpreted in terms of a particular 
type of the top quark mass.  The latter requirement forces us to select kinematic distributions that are infra-red 
safe, so that their computations in higher-orders of QCD perturbation theory can be performed. 
In addition,  methods for measuring the top quark mass should, ideally, be immune 
to contamination from beyond the Standard Model physics --  a scenario that  is conceivable if there is top-like BSM physics 
at the energy scale close to $2 m_t$.  For example, if 
$m_t$ is determined  from the total cross-section $\sigma_{pp \to t \bar t}$ and if the measurement 
of $pp \to t \bar t$ 
receives unknown contributions from top-like BSM physics, 
the extracted value of the top quark mass will be smaller than  the true $m_t$.  This scenario can 
occur for example in SUSY models with  light stop squarks $m_{\tilde t} \sim m_t$  that are still
not excluded experimentally (cf. discussion in Section~\ref{sec:stops}).

\begin{table}[t]
\begin{center}
\begin{tabular}{|c|c|c|c|c|c|c|c|c|}
\hline
& {\rm Ref.}\cite{Chatrchyan:2013boa} & \multicolumn{3}{c|} {\rm Projections}\\
\hline
{\rm CM Energy} & {\rm 7 TeV } & \multicolumn{3}{c|}{\rm 14 TeV  } \\
\hline
{\rm Luminosity} & {$5 fb^{-1}$} & {$100 fb^{-1}$} & {$300 fb^{-1}$} & {$3000 fb^{-1}$} \\ 
\hline\hline
{\rm Syst. (GeV)} & 1.8 & 1.0 & 0.7 & 0.5 \\
\hline
{\rm Stat. (GeV)} & 0.90 & 0.10 & 0.05 & 0.02 \\
\hline\hline
{\bf Total } & {\bf 2.0} & {\bf 1.0} & {\bf 0.7} & {\bf 0.5} \\
\hline
\end{tabular}
\vspace*{0.5cm}
\caption{Projections for the uncertainty in $m_t$ determined using the CMS end-point 
method \cite{Chatrchyan:2013boa}. Extrapolations are based on the published CMS analysis.}
\label{endpoint}
\end{center}
\end{table}

Methods for top quark mass determination that are based on the 
analysis of kinematic distributions of top quark decay products are as close to an ideal method as possible. 
The main reason is that, up to small effects related to selection cuts and combinatorial backgrounds, 
the kinematic variables involved in the analysis can often be chosen to be Lorentz invariant in which 
case they decouple the production stage 
from the decay stage. This minimizes impact of any physics, BSM or SM,  related to $t \bar t$ production  on the top 
quark mass measurement.  Some of these methods are also insensitive to the physics of top quark decay 
and are entirely driven by energy-momentum conservation. 
We will describe two 
of the methods that belong to this category -- the ``end-point'' method developed recently by the CMS 
collaboration \cite{Chatrchyan:2013boa}
and the $``J/\psi''$ method suggested long ago in Ref.~\cite{Kharchilava:1999yj}.

The idea of the end-point method is based on the observation that the invariant mass distribution of a lepton 
and a $b$-jet  contains a relatively 
sharp edge whose position is correlated with $m_t$. Therefore, by measuring the position of this  end-point, 
one can determine the top quark mass.  The number of events close to the end-point 
is fitted to a linear combination of a  flat background and a linear function 
$N_{lb} \sim N_{\rm bck} + S ( m_{\rm lb} - m_{0} ) $, where  $m_0$ gives the position of the end-point. 
The attractive feature of this method is that it is (almost) independent of any assumption about the matrix element 
and that it clearly measures either the pole mass {\it or} some ``kinematic'' mass which is close to it.  
At the small expense of being more model-dependent, one can actually improve on this method by utilizing 
not {\it only} the position of the end-point but also the shape of the $m_{\rm lb}$ distribution. 
Note that away from the kinematic end-point the shape of $m_{lb}$ distribution  is accurately predicted through NLO  QCD including 
off-resonance contributions and  signal-background interferences \cite{Denner:2012yc,Bevilacqua:2010qb}. Close 
to the end-point re-summed predictions are probably required  and are not available at present.

Nevertheless, even without potential    improvements, the 
end-point method offers an interesting alternative to conventional methods. Uncertainties in $m_t$  that 
one may hope to achieve are estimated  in Table~\ref{endpoint}.
We note that by using the end-point method we {\it do  gain in precision  
by going to high-luminosity LHC}. Our projections show that the error as small as $0.5~{\rm GeV}$ can be reached.   
The dominant contributions to systematic uncertainty for 
each of  these studies are the jet-energy scale and hadronization uncertainties. Similar to estimates of 
$\delta m_t$ that can be achieved using conventional methods, we add $300~{\rm MeV}$ to the systematic uncertainty 
in Table~\ref{endpoint}, to account for unforeseen sources of the systematics. 

Another approach to measuring the top quark mass that is very different from conventional ones 
is the so-called $J/\psi$ method~\cite{Kharchilava:1999yj}.  
Here the top quark mass is obtained from fits to the invariant mass distribution of three leptons 
from the {\it exclusive} decays 
of the top quark $t \to e B  \to e J/\psi X \to eee X$, where $X$  denotes light hadrons. 
The extrapolations for the $J/\psi$-method are shown in Table~\ref{JPsi}. 
The attractive feature of this approach is its absolute complementarity to more traditional methods discussed 
above. The uncertainties  in  case of the $J/\psi$ method are dominated by statistical uncertainties for 
luminosities below  $100~{\rm fb}^{-1}$ and by 
theory uncertainties for higher luminosities.  The theory uncertainties in $m_t$ 
 are estimated to be of the order of $1~{\rm GeV}$; they are caused 
by scale and parton distribution functions 
 uncertainties and by  uncertainties in $b \to B$ fragmentation function. Some reduction  of theory 
uncertainties can be expected,  although dramatic improvements in our knowledge of the fragmentation function 
are not very likely.  This is reflected in the change of the theory error shown in Table~\ref{JPsi} 
for $14~{\rm TeV}$ LHC with $3000~{\rm fb}^{-1}$ where it is assumed that NNLO QCD computation 
of the exclusive production of $J/\psi$ in $t \bar t$ events will become available and that the scale 
uncertainty will be reduced by a factor of two. 

\begin{table}[t]
\begin{center}
\begin{tabular}{|c|c|c|c|c|c|c|c|c|}
\hline
& {\rm Ref. analysis} & \multicolumn{5}{|c|}{\rm Projections} \\
\hline
{\rm CM Energy} & {\rm 8 TeV } & \multicolumn{3}{c|}{\rm 14 TeV  } & {\rm 33 TeV} & {\rm 100 TeV}\\
\hline
{\rm Luminosity} & {$20 fb^{-1}$} & {$100 fb^{-1}$} & {$300 fb^{-1}$} & {$3000 fb^{-1}$} & {$3000 fb^{-1}$} & {$3000 fb^{-1}$} \\ 
\hline\hline
{\rm Theory (GeV)} & - & 1.5 & 1.5 & 1.0 & 1.0 & 0.6 \\
\hline
{\rm Stat. (GeV)} & 7.00 & 1.8 & 1.0 & 0.3 & 0.1 & 0.1 \\
\hline
{\bf Total} & {\bf -} & {\bf 2.3} & {\bf 1.8} & {\bf 1.1} & {\bf 1.0} & {\bf 0.6}\\
\hline\hline
\end{tabular}
\vspace*{0.5cm}
\caption{Extrapolations of uncertainties in top quark mass measurements that can be 
obtained with  the $J/\Psi$ method.}
\label{JPsi}
\end{center}
\end{table}

We note that other methods of measuring $m_t$ with relatively high precision are possible and 
were, in fact,  discussed in the literature.  On the experimental side, the three-dimensional 
template fit  method 
was recently presented by the ATLAS collaboration \cite{ATLAS:2013coa}.  The key idea here 
is to determine the top quark mass, the light-quark jet energy scale and the $b$-quark jet energy 
scale from a simultaneous fit to data, thereby transforming a large part of the systematic uncertainty 
related to jet energy scales to a statistical one.  While this measurement determines  the ``Monte Carlo'' 
mass and the  error on this measurement is not competitive 
with other $m_t$-determinations at the moment, its key idea  can be applied in conjunction 
with other methods and will,  hopefully,  help to reduce systematic uncertainties. 
Another potentially interesting 
opportunity is provided by the top quark mass measurements based on 
exploiting $m_t$-dependence of lepton kinematic distributions.
Although such studies were not actively pursued experimentally, they may offer 
an interesting avenue for the top quark mass measurement in the high-pile-up scenario given 
their independence of jet energy scale uncertainties.  Theoretical studies of some lepton distributions 
and their sensitivity to $m_t$ were performed through NLO QCD in Ref.~\cite{Biswas:2010sa} with 
the conclusion that an ${\cal O}(1.5)~{\rm GeV}$ error on $m_t$ can be achieved; further studies 
that include more realistic estimates of uncertainties are clearly warranted.  Finally, it was proposed
recently to employ $t \bar t j$ events to constrain the top quark mass
\cite{Alioli:2013mxa}.  This method is clean theoretically and appears 
to be feasibly experimentally; as shown in Ref.~\cite{Alioli:2013mxa},
an ${\cal O}(1)~{\rm GeV}$ uncertainty  in $m_t$ can be achieved.

The top quark width of 1.4~GeV is too narrow to be measured directly at the LHC. It can be probed
indirectly through single top quark production~\cite{Abazov:2012vd}, which can be determined
to about 5\% at high-luminosity LHC, 
see Section~\ref{sec:topcouplings}. The width can be measured directly to a few percent
through a top pair threshold scan at a lepton collider~\cite{Baer:2013cma,Martinez:2002st}.

We conclude by making a general remark about the future of the top quark measurements at a hadron collider. 
While hadron collider measurements of the top quark mass {\it cannot} compete with $e^+e^-$ colliders,  
our discussion shows that it is possible to have a number of top quark measurements at the LHC, 
including the high-luminosity option, which are clean theoretically  and show high sensitivity to $m_t$. 
It is also important to stress that these measurements are typically limited by different types of uncertainties, so that 
combining their results under the assumption that errors are uncorrelated is a reasonable thing to do.  A combination 
of the results of different measurements, that determine theoretically well-defined top quark mass, 
can lead to further reduction in the error on $m_t$ that 
is achievable at the LHC, 
pushing it into a $0.3 - 0.4~{\rm GeV}$ range.  Further reduction of the uncertainty 
in the top quark mass determination is  possible at suggested $e^+e^-$ machines (ILC, CLIC, TLEP). Such measurements 
are important for testing if the Standard Model {\it without further extensions} can be consistently extrapolated 
to Planckian energy scales. Interest in such studies should increase if no new physics is found at the Run 2 at the LHC.

\section{Top quark couplings} 
\label{sec:topcouplings}
~
The couplings of the top quark to the $W$~and $Z$~bosons, photon, gluon,
and the Higgs boson are explored in this section. It is particularly important to make a direct
measurement of the top quark-Higgs boson Yukawa coupling. Simple estimates suggest that 
typical BSM physics at the TeV scale modifies the top quark couplings to gauge bosons
at the percent level~\cite{Juste:2006sv} but, at the same time, larger 
${\cal O}(10 \%)$ shifts are still possible (see also discussion in Section~\ref{sec:newphysics}). 
Also, our knowledge of the top quark Yukawa 
coupling is poor at the moment and the direct measurement of this coupling with 
any precision is very important. Modifications of top quark couplings typically 
lead to a more complex structure of the interaction vertices, going well beyond simple-minded 
re-scaling of SM couplings. This creates additional complications
and requires us to understand how all the different couplings can be disentangled.

We note  that most of the top quark couplings are measured by comparing observed  {\it rates} of
relevant processes with SM expectations. This puts stringent requirements on theoretical 
predictions and experimental control of systematics, making couplings measurements a difficult 
endeavor at the LHC. 
This section compares the precision reach of couplings measurements 
at low-and high-luminosity LHC to that expected at lepton colliders (mainly ILC and CLIC).
Higher-energy hadron colliders are not expected to improve the measurements much beyond the LHC
sensitivity (except possibly for the $t\bar{t}Z$ coupling)
and are thus not studied here. The muon collider allows for the same studies as done
at the ILC, but with smaller beam-related uncertainties and higher luminosity.
TLEP provides larger data samples than the ILC, though only near the $t\bar{t}$ treshold,
and it has insufficient energy to measure 
Yukawa coupling through direct $t \bar t H$ production though it should be able to reach a
sensitivity of ${\cal O}(30 \%)$ to the $ttH$ coupling from a threshold scan. 
The top quark couplings sensitivity
is compared here using the anomalous coupling notation; a related discussion in terms 
of effective operators can be found in Refs.~\cite{AguilarSaavedra:2008zc,Zhang:2010px}.

\subsection{Strong interaction}
The strong coupling constant of the top quark is fixed in the Standard Model by the requirement of
$SU(3)$ color gauge-invariance. The modifications of this coupling can be expected through radiative
corrections which may introduce  additional structures, such as chromoelectric and chromomagnetic
dipole operators  in the $g t \bar t$ vertex. These modifications occur both in the Standard Model
and in models of new physics.
For example, the Higgs exchange between top quarks modifies the strength of gluon-top quark
interaction in top pair production by ${\cal O}(0.5\%)$ while it does not affect the interaction
of light quarks to gluons.

Strong interactions of the top quark are studied in top quark pair production,
including the $t \bar t +{\rm jets}$ processes, both at the Tevatron and the LHC. A
summary of the current prediction and measurements is shown in Table~\ref{tab:topxs}.
An experimental uncertainty of about $5 \%$  on $\sigma(pp \to t \bar t)$ has been achieved
at the $8~{\rm TeV}$ LHC and it is not expected to significantly improve beyond that during further
LHC operations.  The theory prediction for the total cross-section through NNLO QCD is available 
\cite{Czakon:2013goa,Baernreuther:2012ws,Czakon:2012zr}; it shows a residual scale uncertainty 
of about $3.5\%$, comparable to experimental precision.  Note that, at this level
of precision, electroweak corrections may be important; indeed, as shown in a recent 
update~\cite{Kuhn:2013zoa}, the weak corrections to $t \bar t$ production at the LHC 
are close to $-2.5 \%$.  We conclude that, at a few percent level, there is no indication 
that strong interactions of top quarks are significantly different from that of light quarks.

More exotic types of modifications of top quark strong interactions,  such as chromoelectric $d_t$ 
and chromomagnetic $\mu_t$ 
dipole moments of top quarks, are better constrained from changes in kinematic distributions.
We will discuss this in Section~\ref{sec:kinematics}. Ref.~\cite{Baumgart:2012ay} finds that
constraints of 1\% or below are possible with
$100~\mathrm{fb}^{-1}$ at 13~TeV.

Exchanges of axigluons or Kaluza-Klein excitations of gluons not only modify couplings 
of top quarks to gluons, but also  generate 
four-fermion operators that involve light and heavy quarks 
$ \left ( \bar q T^{a} q \right )  \; \left ( \bar t T^{a} t \right )$. 
These operators can be directly probed at the LHC, where the sensitivity to scales 
between $1.2~{\rm TeV}$ and $3~{\rm TeV}$ can be expected~\cite{TopCouplWhitePaper2}.

Finally, top quark coupling to gluons can be probed at a linear collider through a threshold scan. 
The peak cross-section 
at threshold is proportional to $\sigma_{\rm peak} \sim \alpha_s^3/(m_t \Gamma_t)$.   Using the total 
cross-section and other measurements at threshold, one can determine the strong coupling 
constant with better than one percent precision and the total width of the top quark $\Gamma_t$ 
with the precision of a few percent~\cite{Martinez:2002st,Seidel:2013sqa}. 
~
\begin{table}[!h!tbp]
\begin{center}
\begin{tabular}{|l|c|c||c|c|}
\hline
                 &\multicolumn{2}{c||}{Theory prediction}& \multicolumn{2}{c|}{LHC Measurement} \\
CM Energy [TeV]  &  7 & 8             &  7  &   8      \\
Luminosity [fb$^{-1}$]& &             & 1-5 & 2-15      \\ 
\hline
Top pairs $\sigma(t\bar{t})$ [pb]& $172\pm7$~\cite{Czakon:2013goa}&$246\pm10$~\cite{Czakon:2013goa}&$173\pm10$&$241\pm32$ (ATLAS)~\cite{ATLAS-CONF-2012-149} \\
         & & &(LHC comb.)~\cite{ATLAS-CONF-2012-134}   &$227\pm15$ (CMS)~\cite{CMS-PAS-TOP-12-007} \\
Single top $\sigma$(t-chan) [pb]& $66\pm2$~\cite{Kidonakis:2012rm} & $87\pm3$~\cite{Kidonakis:2012rm} &$83\pm20$ (ATLAS)~\cite{Aad:2012ux}& $95\pm18$ (ATLAS)~\cite{ATLAS-CONF-2012-132} \\
          & & &$67\pm6$ (CMS)~\cite{Chatrchyan:2012ep}&$80\pm13$ (CMS)~\cite{CMS-PAS-TOP-12-011} \\
Single top $\sigma(Wt)$ [pb]&$15.6\pm1.2$~\cite{Kidonakis:2012rm}& $22.2\pm1.5$~\cite{Kidonakis:2012rm}& $16.8\pm5.7$ (ATLAS)~\cite{Aad:2012xca}& $27.2\pm5.8$(ATLAS)~\cite{ATLAS-CONF-2013-100} \\
          & & &$16\pm4$ (CMS)~\cite{Chatrchyan:2012zca}& $23.4\pm5.4$ (CMS)~\cite{CMS-PAS-TOP-12-040}\\
\hline
\end{tabular}
\vspace*{0.5cm}
\caption{LHC single top and top pair production cross-section measurements.
}
\label{tab:topxs}
\end{center}
\end{table}

\subsection{Weak interactions: W boson}
~
The coupling of the top quark to the $W$~boson is studied in top quark decays
and in single top quark production at the LHC and the Tevatron, and in top quark 
decays at the linear collider. The effective Lagrangian describing the $Wtb$ 
interaction including operators up to dimension five is~\cite{AguilarSaavedra:2008zc}
\begin{eqnarray}
\mathcal{L}&=&-\frac{g}{\sqrt{2}}\bar{b} \gamma^{\mu}
(V_L P_L + V_R P_R) t W_{\mu}^{-}
-\frac{g}{\sqrt{2}} \bar{b} \frac{i\sigma^{\mu\nu} q_{\nu}}{M_W} (g_L P_L + g_R P_R) t W_{\mu}^{-} 
+ h.c. \, ,
\label{eq:Wtbcoupling}
\end{eqnarray}
 where  $M_W$ is the mass of the $W$~boson, $q_{\nu}$ is its four-momentum, 
$P_{L,R}=(1 \mp \gamma_5)/2$ are the left- (right-) handed projection operators, and  $V_L$
is the left-handed coupling, which in the SM is equal to the 
Cabibbo-Kobayashi-Maskawa matrix element $V_{tb}$~\cite{Cabibbo:1963yz}. 
 The right-handed vector coupling $V_R$ and the
left-and right-handed tensor couplings $g_L$ and $g_R$ may only appear in the SM through radiative
corrections. 

The measurement of helicity fractions of $W$~bosons through lepton angular distributions
in top quark decays can distinguish SM-like left-handed vector couplings from right-handed vector
and from left-or right-handed tensor couplings. With the data 
collected at the 8~TeV LHC, $V_R, g_L$ and $g_R$ can be constrained to be smaller than $0.1$.
We note that theoretical predictions for $W$-boson helicity fractions in the SM  have been 
extended to NNLO QCD \cite{Czarnecki:2010gb,Gao:2012ja,Brucherseifer:2013iv} and, therefore, theory
uncertainties on helicity fractions are about one order of magnitude smaller than experimental
one. Measuring the helicity fraction to a similar level at the high-luminosity LHC and beyond
is therefore necessary to obtain the best sensitivity to new physics.

Single top quark production involves the $tWb$ vertex in top quark production and
thus also provides information on the magnitude of the $tWb$ coupling and the 
CKM matrix element $|V_{tb}|$. Single top quarks are produced in three different modes: the
``$t$-channel'' mode where a $W$~boson is exchanged between a light quark line and a heavy
quark line,  which has the largest cross-section;  the ``$Wt$ associated production'' mode where
either the decay or the exchange of a virtual $b$~quark leads to the final state of a top quark
and a $W$~boson, with the next-to-largest cross-section; and the ``$s$-channel'' production 
and decay of a virtual $W$~boson, which has a very small cross-section. The LHC
cross-section measurements for $t$-channel and $Wt$ together with the corresponding prediction
are shown in Table~\ref{tab:topxs}.
The three modes have different sensitivities to new physics and anomalous couplings.
LHC measurements of single top quark production, in particular in the $t$-channel
mode,  are also sensitive to off-diagonal CKM matrix elements~\cite{Lacker:2012ek}. The
single top production cross-section measurement already is dominated by systematic
uncertainties~\cite{Aad:2012ux,Chatrchyan:2012ep,Aad:2012xca}, and
the situation is not expected to improve much at higher energies or with larger
datasets. The ultimate cross-section uncertainty will likely be around 5\%, similar to
top pair production, so that uncertainties on $tWb$ coupling  and $|V_{tb}|$ will be close to
2.5\%~\cite{TchanWhitePaper}. Searches for anomalous couplings in the $tWb$ vertex depend on the
ability to separate the signal from both SM single top and from large backgrounds and  are less
limited by systematic uncertainties.
A search for CP violation through an anomalous coupling gives a limit on
$Im(g_R)$~\cite{ATLAS-CONF-2013-032}. Finally, an extrapolation of the sensitivity to anomalous
couplings from single top quark production and decay shows that with 300~fb$^{-1}$  the anomalous
couplings as small as $0.01$ can be probed.

Electron-positron colliders are expected to do a comparable job in exploring the strength of 
$tWb$ interaction vertex by considering the cross-section scan  of 
$\sigma_{t b W}$ cross-section  at CM energies between $m_t$ and $2m_t$. It was estimated 
in Ref.~\cite{Juste:2006sv} that $g_{tWb}$ can be measured with the precision of about two percent. 
Among more  exotic options is the possibility to study $tWb$ interaction 
at a $\gamma e$ collider, with a reach of
$10^{-1}$ to $10^{-2}$~\cite{Boos:2001sj}. The reach is about $10^{-3}$ to
$10^{-2}$ for a LHC-based electron-proton collider with a CM energy of
1.3~TeV~\cite{Dutta:2013mva}.

Knowledge of $tWb$ interaction can be used to compute the top quark decay width $\Gamma_t$.
This can be compared to direct measurements of $\Gamma_t$, discussed in Section~\ref{sec:topmass}.

\subsection{Electroweak interaction: Z boson and photon}
The interaction of the top quark with neutral electroweak gauge bosons has not been studied in 
detail so far. Indeed, although both the charge of the top quark~\cite{Aad:2013uza}
and the production cross-section of top pair in association with a photon were measured
experimentally~\cite{ATLAS-CONF-2011-153}, this does not give us all the information
required to fully constrain the $t \bar t \gamma $ vertex.  The interaction of top quarks with 
the $Z$~boson has not been measured yet. Similarly to other coupling, a measurement with
${\cal O}(10\%)$ precision will 
be useful for constraining models of physics beyond the Standard Model. As an example,
Section~\ref{sec:newphysics} discusses compositeness, which would be constrained by a measurement
at this precision.

It is challenging, but perhaps not impossible,  to probe  $t\bar t Z$ and $t \bar t \gamma$
couplings at the LHC with 10\% precision. A lepton collider would provide even higher precision.

A general expression for $t \bar t V$, $V=\gamma, Z$ interaction vertex is~\cite{Juste:2006sv}
\begin{equation}
\Gamma_\mu^{ttX}=ie\left\{-\gamma_{\mu}\left((F_{1V}^X+F_{2V}^X) + \gamma_5 F_{1A}^X\right)
            + \frac{(q-\overline{q})_\mu}{2m_t}\left(
               F_{2V}^X-i\gamma_5 F_{2A}^X\right) \right\}\,,
\label{eq:tZg}
\end{equation}
where $X$ is either a photon ($X=\gamma$) or $Z$~boson ($X=Z$). The couplings 
$F_{1V}^\gamma$, $F_{1V}^Z$ and $F_{1A}^Z$ have tree-level SM values. 

The LHC experiments have measured the production of photons in association with
top quark pairs, and will measure both the $\gamma+t\overline{t}$ and
$Z+t\overline{t}$ cross-sections. However, in both cases, significant kinematic 
cuts on final state particles are required either to suppress the backgrounds or, in case 
of photons, to select events where photons are emitted from top quarks rather than from their 
decay products \cite{Baur:2005wi,Baur:2004uw,Juste:2006sv}.
Therefore, extracting the top-photon or top-$Z$ coupling from the associated production
is difficult;
it relies on a detailed theoretical understanding of the production
process. This is becoming available thanks to recent studies of 
$pp \to t \bar t \gamma$ and $pp \to t \bar t Z$ processes  in next-to-leading order in 
QCD~\cite{Lazopoulos:2008de,Melnikov:2011ta,Garzelli:2012bn,Garzelli:2011is}.
Single top quark production in association with a $Z$~boson can also
be used to study the $tZ$ coupling~\cite{Campbell:2013yla}. 

Measurements of the $t \bar t \gamma$ and $t \bar t Z$ couplings with the 
highest precision can be performed at a linear collider~\cite{Baer:2013cma}.
The two couplings are entangled in the top pair production process. Separating the two couplings
requires polarized beams. For the projections
in Table~\ref{tab:topZxs}, electron and positron polarizations of 80\% and 30\%,
respectively, are assumed.  It follows from Table~\ref{tab:topZxs} that 
most of the top quark couplings to the photon and the $Z$~boson can be measured
at a linear collider (ILC/CLIC) to a precision that is typically an order of magnitude better than
at the LHC. Despite the lack of detailed studies,
the precision on the combined coupling accessible at TLEP is expected to be even better than
that at the linear collider due to the higher integrated luminosity. However, the lower energy
and lack of beam polarization make it impossible to disentangle the $\gamma$ and $Z$~couplings
and the different couplings in Eq.~\ref{eq:tZg}. Similarly,
detailed studies also have not been performed for a muon collider, which provides larger
integrated luminosity and smaller beam uncertainties but also
challenging backgrounds; thus it is not clear if it will be able to improve on the linear collider
measurements.

In summary, although a linear collider will achieve the highest precision in the $t \bar t Z$ 
and $t \bar t \gamma$ coupling measurements, it is clear that the LHC -- and in particular its 
high-luminosity 
phase -- will be able to probe these  couplings in an interesting precision range where deviations
due to generic BSM physics are expected.

~
\begin{table}[!h!tbp]
\begin{center}
\begin{tabular}{|l|c|c|c|c|c|c|c|c|}
\hline
Collider         & \multicolumn{2}{c|}{LHC}& ILC/CLIC \\ 
CM Energy [TeV]          &  14  &   14  & 0.5     \\
Luminosity [fb$^{-1}$]   & 300  & 3000  & 500     \\ 
\hline
SM Couplings             &      &       & \\
~~photon, $F_{1V}^\gamma$ (0.666) & 0.042& 0.014 & 0.002 \\
~~$Z$ boson, $F_{1V}^Z$  ( 0.24) & 0.50 & 0.17  & 0.003 \\
~~$Z$ boson, $F_{1A}^Z$  (0.6) & 0.058& --  & 0.005 \\
Non-SM couplings         &      &       & \\
~~photon, $F_{1A}^\gamma$ & 0.05 & --  & -- \\
~~photon, $F_{2V}^\gamma$ & 0.037& 0.025 & 0.003 \\
~~photon, $F_{2A}^\gamma$  & 0.017& 0.011 & 0.007 \\
~~$Z$ boson, $F_{2V}^Z$ & 0.25 & 0.17  & 0.006 \\
~~$Z$ boson, $ReF_{2A}^Z  $& 0.35 & 0.25  & 0.008 \\
~~$Z$ boson, $ImF_{2A}^Z   $& 0.035& 0.025 & 0.015 \\
\hline
\end{tabular}
\vspace*{0.5cm}
\caption{Expected  precision of the 
top quark coupling measurements to the photon and the $Z$~boson at the
LHC~\cite{TopCouplWhitePaper1,TopCouplWhitePaper2} and the linear
collider~\cite{Baer:2013cma}.  Expected magnitude of such couplings 
in the SM is shown in brackets. Note that   the ``non-standard model'' couplings appear 
in the Standard Model through radiative corrections; their expected magnitude, therefore, 
is $10^{-2}$. 
}
\label{tab:topZxs}
\end{center}
\end{table}

\subsection{Yukawa coupling}
The coupling of the top quark to the Higgs boson is of great interest. Since the
top quark provides one of the largest contributions to the mass shift of the
Higgs boson, any deviation in the $t\bar{t}H$ coupling from its Standard Model value may have
far-reaching consequences for the naturalness problem.
The coupling of the top quark to the Higgs boson can be  measured at the LHC
in different  final states. It will also be studied in detail at lepton colliders. More
details on the top Yukawa coupling measurements can be found in the Higgs working
group report~\cite{HiggsWhitepaper}.

The process $pp \to t \bar t H$ can be studied in a variety of final states,
depending on the top quark decay mode (lepton+jets or dilepton or all-jets) and the Higgs decay
mode ($b\overline{b}$, $\gamma\gamma$,  $WW$ {\rm etc.}).
Each final state has a its own, typically large, background, mainly from top
quark pair production in association with jets or electroweak bosons. The coupling of the top
quark to the Higgs boson is extracted from these measurements with relatively large
uncertainties of about 20\% initially, with an improvement to 10\% at the high-luminosity 
LHC~\cite{PhysRevD.86.073009,Onyisi:2013gta,CMS-NOTE-2012-006,ATLAS-collaboration:1484890,TopCouplWhitePaper1,TopCouplWhitePaper2}.
At the high-luminosity LHC, the $t\overline{t}H$ final state is also a promising channel 
to measure the muon coupling of the Higgs boson~\cite{TtHmumuWhitePaper}, though it is still
statistics-limited. Production of $t\overline{t}H$ with Higgs decay to photons is observable
at the LHC~\cite{TopCouplWhitePaper2}, which allows for a study of the CP structure of the
top-Higgs vertex~\cite{Gunion:1996xu}.

Better precision in the top-Higgs coupling can be achieved at  lepton colliders running
at a sufficiently high CM energy and collecting large integrated luminosity.
Initial studies focused on a CM energy of 800~GeV where the $t\overline{t}H$ cross-section
 is largest, however a measurement at 500~GeV is also possible. For the
projections in Tab.~\ref{tab:topHiggs}, electron and positron polarizations of 80\% and
30\%, respectively, are used. For the ILC/CLIC, the nominal luminosity for 1~TeV running
is assumed, which corresponds to twice the ILC luminosity at 500~GeV.
A comparison of the top Yukawa coupling precision expected at different colliders is
shown in Table~\ref{tab:topHiggs}, from where it follows that a linear collider provides
improvements compared to the high-luminosity LHC. It is also possible to
measure the Yukawa coupling in a threshold scan that is sensitive to the modification of the
$t \bar t$ production cross-section through a Higgs exchange. A precision
of ${\cal O}(30)\%$ can, perhaps, be achieved in this case. Note that this is the only way 
to get information on the top Yukawa coupling at TLEP.

\begin{table}[!h!tbp]
\begin{center}
\begin{tabular}{|l|c|c|c|c|c|c|c|c|c|}
\hline
Collider         & \multicolumn{2}{c|}{LHC}& ILC & ILC  & CLIC \\ 
CM Energy [TeV]          &  14  &   14  & 0.5    & 1.0  & 1.4  \\
Luminosity [fb$^{-1}$]   & 300  & 3000  & 1000   & 1000 & 1500 \\ 
\hline
Top Yukawa coupling $\kappa_t$& ($14-15$)\%& ($7-10$)\% & 10\% & 4\% & 4\%\\
\hline
\end{tabular}
\vspace*{0.5cm}
\caption{Expected  precision of the top quark Yukawa coupling measurement
expected at the LHC and the linear collider~\cite{HiggsWhitepaper}. The range
for the LHC precision corresponds to an optimistic scenario where systematic
uncertainties are scaled by 1/2 and a conservative
scenario where systematic uncertainties remain at the 2013
level~\cite{CMS-NOTE-2012-006,ATLAS-collaboration:1484890,Onyisi:2013gta}.
The ILC~\cite{Baer:2013cma,Yonamine:2011jg}
and CLIC~\cite{Abramowicz:2013tzc} projections assume polarized beams and nominal integrated
luminosities.
}
\label{tab:topHiggs}
\end{center}
\end{table}

\section{Kinematics of top-like final states} 
\label{sec:kinematics}
~
Working with top quarks requires us to understand how they are produced 
and how they decay. In this section, we discuss what we know about that and what we can learn 
in the future.  While such a discussion is interesting in its own right, it also allows us to understand 
to what extent deviations from expected behavior of various top quark distributions 
in different kinematic regimes can be probed at existing and future facilities.  In general, after the run I
of the LHC and the studies of top quark pair production at the Tevatron, it is fair to say that
dynamics of $t \bar t$ production is well-understood. The only, but significant,
discrepancy that exists is the disagreement between forward-backward asymmetry for top quarks
expected in the Standard Model and the measured value of this asymmetry at the Tevatron. Is it
possible to clarify the situation with forward-backward asymmetry at the LHC or other future
facilities? This is a data-motivated question that we address in this section.

\subsection{Kinematic distributions in top quark pair production}
Our current understanding of top quark pair production in hadron collisions is based 
on next-to-leading order computations for the fully-differential process 
$pp \to t \bar t \to W^+W^- b \bar b$ both within and beyond the narrow width 
approximation~\cite{Denner:2012yc,Bevilacqua:2010qb,Melnikov:2009dn,Campbell:2012uf}. 
The comparison of these 
computations ensures that the narrow width approximation works very well at the LHC unless one
moves to extreme kinematic regimes where production of two on-shell top quarks becomes
kinematically unfavorable. The success of  the narrow width approximation in $t \bar t$ production
allows us to claim its validity for more complicated processes, such us production of top quark pairs
in association with jets \cite{Melnikov:2010iu,Dittmaier:2008uj,Kardos:2011qa} 
or with gauge bosons, that we will discuss in the next Section. 
Existing theoretical results on top quark pair production 
will be further improved by extending available results for differential 
quantities to next-to-next-to-leading order  in perturbative QCD. We note that 
 such results for  the total cross-section $pp \to t \bar t $ were recently obtained
\cite{Czakon:2013goa,Baernreuther:2012ws,Czakon:2012zr}.  

\begin{figure}[!h!tbp]
  \centering
  \label{fig:topkin-basicdistr}
\includegraphics[width=0.7\hsize]{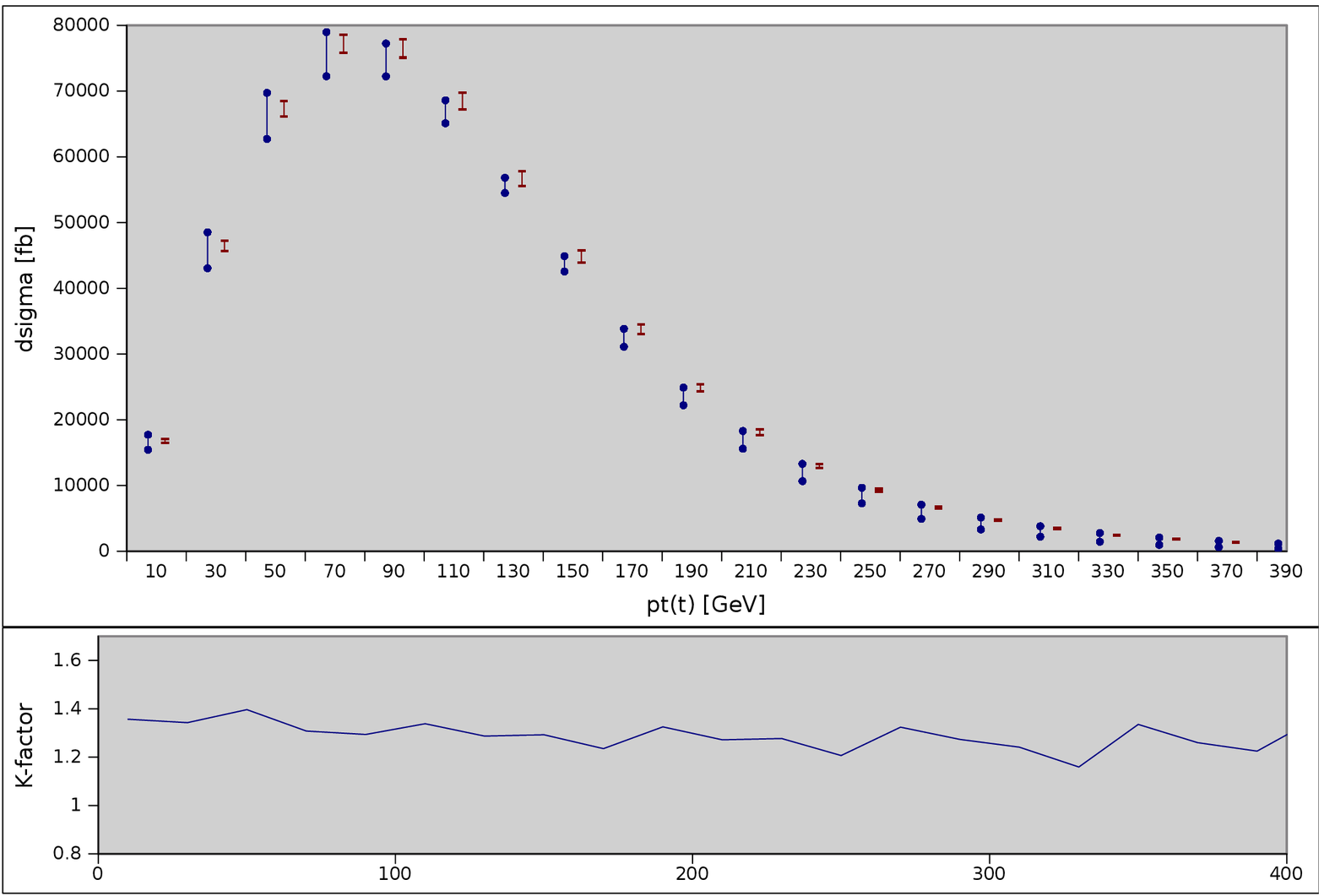}
\includegraphics[width=0.7\hsize]{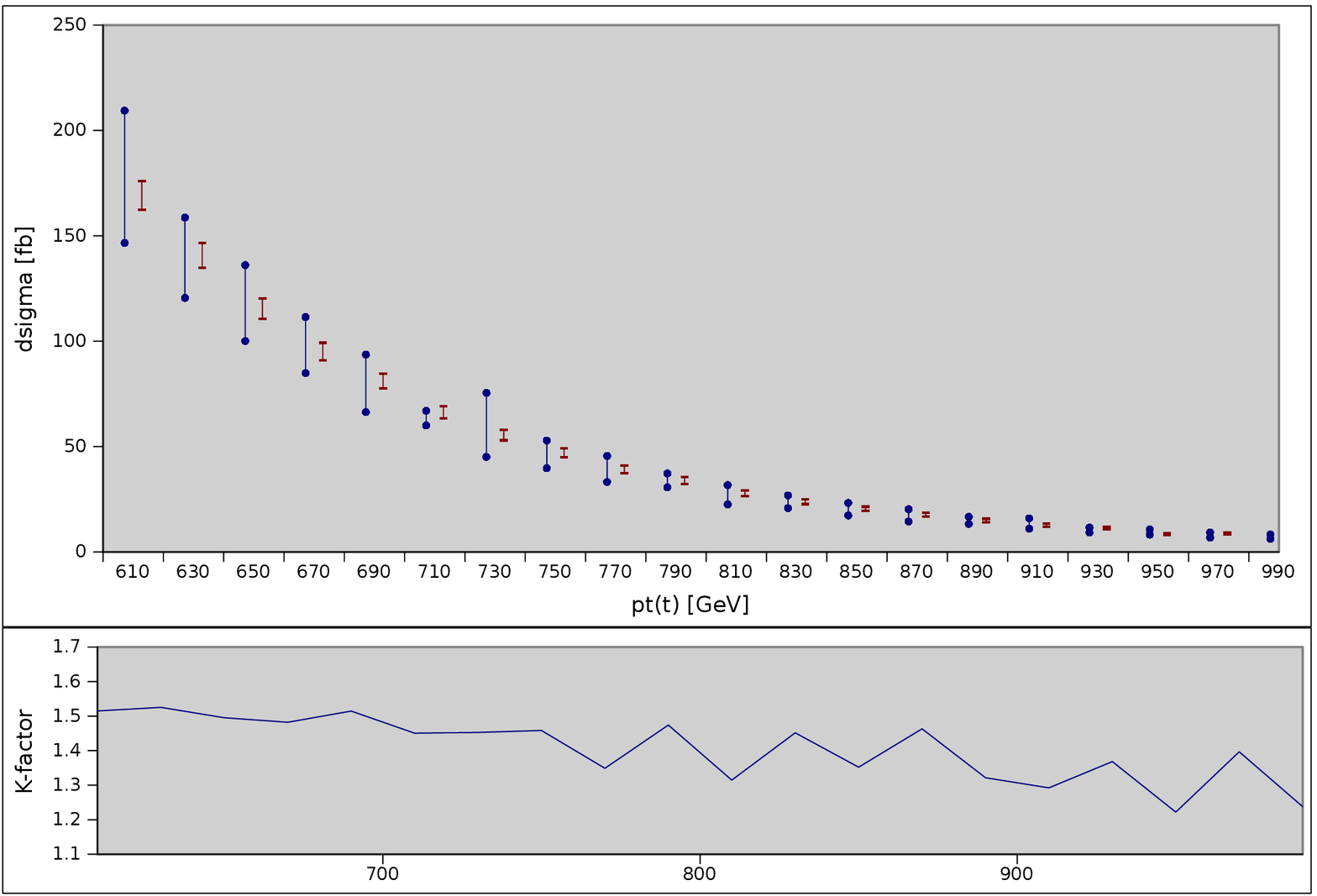}
  \caption{NLO QCD predictions \cite{Campbell:2010ff} for the transverse momentum of the top
quark at the $14$~TeV LHC.
Blue error bars correspond to the central MSTW pdf set and scale variation by a factor of two around $\mu = m_t$. 
Dark red error bands correspond to 1 SD variation of MSTW pdf error sets for fixed renormalization and factorization scale 
at $\mu = m_t$. Note that the red and blue bars can be off-set because at NLO the central scale does not necessarily corresponds to the center of the blue bar. In this case, it seems that it is towards the upper value of the blue bar.
}
\end{figure}

We will now take a closer look at the quality of theoretical description of various kinematic
distributions. To this end, we show distributions in the  top quark transverse momentum 
$p_\perp$ in $pp \to t \bar t$ at  the 14~TeV LHC in Fig.~\ref{fig:topkin-basicdistr}
and indicate the uncertainties in the  predictions caused by imperfect knowledge of parton
distribution functions and missing higher-order corrections. We estimate the latter by varying 
renormalization and factorization scales by a factor of two around the fixed value $\mu = m_t$. 
The computations are performed with MCFM~\cite{Campbell:2010ff}.
We 
see that scale  uncertainties dominate and that uncertainties in  theory predictions 
 are at the level of 20\%. 

Another interesting kinematic regime is when each top quark is produced with a high Lorentz boost (>2), resulting in collimated
decay products that may be clustered into a single jet (``boosted'' topology). As we will see, it is
more difficult to understand the uncertainty in the theoretical prediction for this quantity.
Indeed, a MCFM-based computation shows that for $p_\perp > 800~{\rm GeV}$, the uncertainties on rapidity
and $p_\perp$ distributions roughly double compared to the non-boosted
regime~\cite{TopKine:Snowmass}.  However, these uncertainties may be underestimated. 
Indeed,  resummation computations, either traditional or Soft-Collinear Effective Field Theory (SCET). 
point towards additional positive contributions to $p_\perp$ distributions
at high values of the top quark momentum \cite{Ahrens:2011mw,Auerbach:2013by}.
Forthcoming NNLO computations will be required to resolve this issue.

All kinematic distributions in top quark pair production are routinely checked 
for signs of new physics. Prominent among them is the distribution in the invariant mass 
of a $t \bar t$ pair, which may be significantly modified by the presence of resonances that decay 
to top pairs. Theoretical predictions for such distributions exist both in fixed order  
QCD and in SCET~\cite{Ahrens:2011mw}; they show theoretical errors between ten and fifteen percent, 
depending  on  $m_{t \bar t}$ {\it and}. Similarly to the $p_\perp$ distribution, these show significant differences 
between fixed order and resummed results at large values of $m_{t \bar t}$. 

Other kinematic distributions, such as angular correlations  between either top quarks 
or their decay products, did not lead to  studies at the Tevatron because 
of low statistics. However, such studies at the LHC will become increasingly important 
as the tool to analyze various subtle features  of top quark interactions with with both SM
and, hopefully, BSM particles.  In the following  
subsections, we discuss examples of this, considering top quark spin correlations and the forward-backward $t \bar t$ asymmetry.

\subsection{Top quark spin correlations}
\label{sec:topkin-spincorrel}
~
Spin correlations between $t$ and $ \bar t$ are an interesting feature of 
top quark physics,  related to the fact that top quark lifetime
is so short that $t(\bar t)$ spin information is transferred to their decay products without being affected by non-perturbative 
hadronization effects. 
Observable spin correlations are affected by the structure of $g \bar t t$ 
and  $tWb$ interaction vertices.
After the observation of top quark spin correlations at the
Tevatron \cite{Abazov:2011gi} and recently at the LHC
\cite{ATLAS:2012ao,CMS-PAS-TOP-12-004}, experimental analyses will soon
be able to probe spin correlations in detail. Perhaps, they will use spin correlations 
as an analysis tool to find and constrain physics beyond the Standard Model.

The cleanest $t\bar t$ samples for the study of spin correlations are the ones with two opposite-sign
leptons in the final state. Spin correlations in this dilepton
mode manifest themselves most prominently in the distribution
of the relative azimuthal angle between the two leptons \cite{Mahlon:2010gw}. 
This distribution is robust under higher order corrections and parton showering
effects~\cite{Melnikov:2009dn,Bernreuther:2010ny,Frixione:2007zp}. For standard
acceptance cuts, NLO QCD effects introduce shape changes of at most 20\%. 
If additional cuts are applied that enhance spin
correlations, NLO corrections increase the correlation even further.
Electroweak corrections have negligible effect and scale variations
are small because distributions are typically normalized.
On the experimental side, the
reconstruction of the lepton opening angle in the laboratory frame
is straightforward  and can be done  with small systematic uncertainties.  The normalized
azimuthal opening angle distribution is therefore an ideal observable
for studying top quark spin correlations.  Other
observables such as helicity angles, double differential
distributions and asymmetries can also be explored.

\begin{figure}[!h!tbp]
  \centering
\includegraphics[width=0.7\hsize]{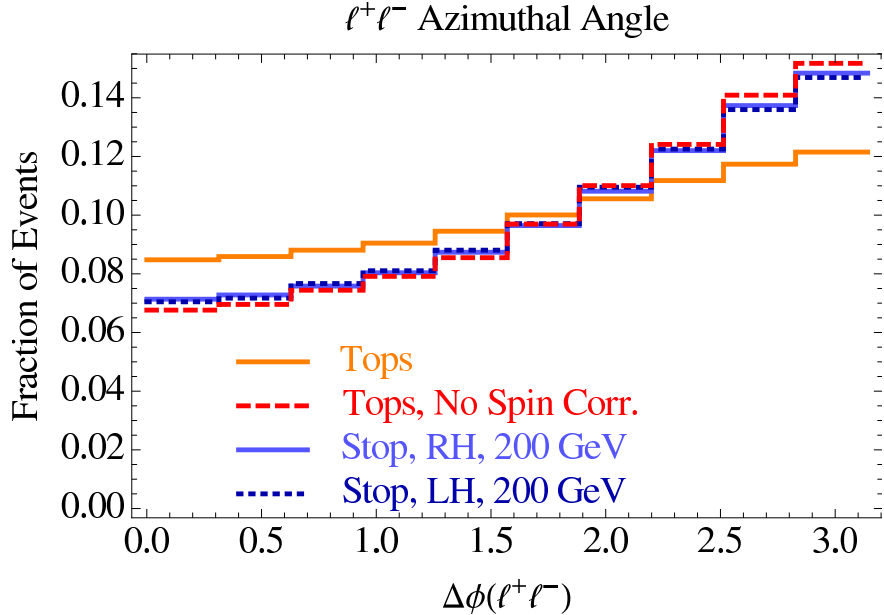}
  \caption{Top quark spin correlation angle for top quark production in the SM and without
spin correlation and for stop quark production with different couplings~\cite{TopKine:Snowmass}.
}
\label{fig:spincor}
\end{figure}

The utility of top quark spin correlations to search for physics beyond the Standard Model 
stems from the vector coupling of top quarks 
to gluons, from the fermion nature of the top quark, and from the couplings of a top quark to 
a $W$~boson and a $b$~quark through a left-handed vector current.  Any changes in that 
list must lead to an observable change in the spin correlation pattern.
For example, it has been shown that top quark spin correlations can be used to
distinguish SM top quarks from scalar partners (stops) even if tops and stop 
are degenerate in mass \cite{Han:2012fw}. The potential of spin correlations 
to distinguish   SM top pair production and  
stop ( $m_{\tilde t} = 200~{\rm GeV}$ ) pair production is illustrated  in 
Fig.~\ref{fig:spincor}~\cite{TopKine:Snowmass}.
~

Modifications of the $g \bar t t$ vertex, that can be parametrized in 
terms of   top quark chromomagnetic
$\hat\mu_t$ and electric $\hat d_t$ dipole moments, can be exposed through 
spin correlations in the dileptonic and in the semileptonic
channels~\cite{Baumgart:2012ay,Bernreuther:2013aga}. 
Here, the magnetic and electric dipole moments $\hat \mu_t$ and $\hat d_t$ correspond to the Lagrangian
 $\frac{1}{2} G^a_{\mu\nu} \left( \tilde\mu_t \bar t \sigma^{\mu\nu} T^a t + \tilde d_t \bar t i \sigma^{\mu\nu} \gamma_5 t \right)$
where we write
 $ \tilde \mu_t = \frac{g_s}{m_t} \hat \mu_t$ and $\tilde d_t = \frac{g_s}{m_t} \hat d_t$. 
Indeed, using dilepton events sample of the $20~\mathrm{fb}^{-1}$
run at 8~TeV, it should be possible to constrain
$\mathrm{Re}(\hat\mu_t)$ and $\mathrm{Re}(\hat d_t)$ at the few percent
level.  The imaginary parts $\mathrm{Im}(\hat\mu_t)$ and
$\mathrm{Im}(\hat d_t)$ can be constrained with 15-20\% precision from lepton-top helicity
angles in the semileptonic channel where a full reconstruction of the
$t\bar t$ system is possible,  using the same dataset. 
Ref.~\cite{Baumgart:2012ay} finds that constraints 
at the level of one percent or even below are possible with
$100~\mathrm{fb}^{-1}$ at 13~TeV.  Finally, in case of the discovery of a new
resonances which decays into $t\bar t$ pairs, top quark spin
correlations can also be used to analyze couplings of this new
particle~\cite{Baumgart:2011wk,Caola:2012rs}.

\subsection{Top quark pair forward-backward asymmetry}

Top quark pair production in $q \bar q$  collisions exhibits forward-backward asymmetry that
arises in higher orders in perturbative QCD~\cite{Kuhn:1998jr,Kuhn:1998kw,
PhysRevD.78.014008,PhysRevD.77.014003,PhysRevD.73.014008}. As the result, the top quark is
preferentially emitted in the direction of the incoming
quark, while the anti-top quark follows the direction of the incoming antiquark. At the Tevatron, 
the direction of the incoming quark corresponds to the direction of the incoming proton, while 
the incoming anti-quark most likely comes from an anti-proton.  Since LHC is a proton-proton collider,  the
$t \bar t$ asymmetry observation becomes difficult because  directions of quark and anti-quark 
are not correlated with directions of initial hadrons and, in addition, there is a large 
gluon flux that reduces the asymmetry. The forward-backward asymmetry is measured at the LHC through 
the difference in rapidity distributions of $t$ and $\bar t$. The harder spectrum of valence quarks in the 
proton and the correlation of top quark direction with the direction of the incoming quark make the top 
rapidity distribution 
broader than the rapidity distribution of the anti-top. The corresponding asymmetry is referred to as the charge 
asymmetry. It can be written as 
\begin{equation}
A_{C}^{\eta} = \frac{N(\Delta|\eta| > 0)-N(\Delta|\eta| < 0)}{N(\Delta|\eta| > 0)+N(\Delta|\eta| < 0)}
\end{equation}
where $\Delta|\eta|\equiv |\eta_t |-|\eta_{\bar t}| $ tells us  whether
the reconstructed top or anti-top is more central according to
lab-frame {\em pseudo-rapidity}.  

Inclusive  forward-backward asymmetries measured at the Tevatron exceed SM predictions by almost three standard 
deviations~\cite{PhysRevD.87.092002,PhysRevD.87.011103}, with stronger dependence on $t \bar t$ invariant  mass and rapidity 
than predicted by the SM. 
At the LHC,  the ATLAS and CMS Collaborations have performed measurements of 
the charge asymmetry 
$A_{C}$~\cite{ATLAS:2012an, Chatrchyan201228} and found agreement  with SM predictions. However, these measurements 
have large errors that makes them not conclusive.  

Given that the forward-backward asymmetry is the {\it only} measurement in top physics that shows profound disagreement 
with the Standard Model prediction, we feel it is important  to understand wether this problem  can 
be resolved. Our estimates for the LHC are presented below.  At a linear collider, it is not possible to address 
this problem directly unless the asymmetry mediator is light and can be directly studied  in $e^+e^- \to t \bar t jj$.

The higher energy of the $14~{\rm TeV}$ LHC 
increases the fraction  of $t\bar t$ events that arise from gluon
fusion, relative to $7$ and $8$~TeV LHC.  Since $gg \to t \bar t$ does not 
produce an asymmetry, 
the asymmetric signal decreases with increased  center of
mass energy of the collider. Already at $7$~TeV LHC  measurements of the top charge asymmetry
are limited by systematic uncertainties. The situation will not change at a higher-energy machine, but 
higher luminosity can eventually allow us to improve the systematic errors.

SM predictions for the $14$~{\rm TeV} LHC  as a function of cuts on the minimum invariant mass of the top pair
$m_{t\bar t}$ 
are calculated in Ref.~\cite{PhysRevD.86.034026}.  
Cutting on either $t \bar t$ invariant mass
or center-of-mass rapidity increases the proportion of $q \bar q$-initiated top
pair events relative to gluon-initiated events, and thus enhances the
signal.  However,  even with kinematic cuts, the
size of the signal at the 14 TeV LHC is comparable to the systematic
uncertainties on the current measurements.
The dominant contributions to the systematic errors are 
jet energy scale, lepton identification, 
background modeling ($t\bar t$, $W+$ jets, multijets), the model dependence of signal generation
and the unfolding procedure. Several  contributions to systematic errors, such as jet energy scale
and lepton identification, can be reduced  with increased  luminosity.  Possible improvements in
background modeling are less clear.  
The dilepton channel can also be used, usually by defining
a lepton-based asymmetry rather than the top quark based $A_{C}$, with a sensitivity similar to
the lepton+jets one~\cite{ATLAS-CONF-2012-057, CMS-PAS-TOP-12-010}. 

Our estimates of the ultimate LHC sensitivity~\cite{TopKine:Snowmass} show 
that, with sufficient luminosity, the 14~TEV LHC will be able to {\it conclusively} measure 
the SM  asymmetry provided  that largest systematic errors 
(background modeling ($40\%$), lepton identification ($30\%$) and $W+{\rm jets}$ modeling ($13\%$)~\cite{Chatrchyan201228}) 
 scale with luminosity. If the asymmetry is enhanced
due to BSM effects -- as indicated by the Tevatron 
data -- the prospects for observing the asymmetry by CMS and ATLAS become event brighter. 

An internal study of the LHCb collaboration \cite{lhcb} concludes that a 
measurement of the SM  $t \bar t$  asymmetry by LHCb experiment  is possible 
at the $14~{\rm TeV}$ LHC with sufficient luminosity, as suggested earlier in~\cite{Kagan:2011yx}.
This will provide a measurement of $A_c$ at the LHC which is complementary to the  measurement of $A_c$  by ATLAS and CMS  
collaborations. Combining all of the measurements, one can probably achieve a significant improvement 
in the precision compared to individual experiments and hopefully solve the 
forward-backward asymmetry puzzle.

To this end, note that 
out of the vast zoology of proposed BSM explanations for the Tevatron
anomaly in the top forward-backward asymmetry, axigluons
\cite{Frampton:2009rk, Bai:2011ed, Tavares:2011zg} 
are left looking least constrained after the low-energy LHC run has been completed. 
Detailed discussions of experimental constraints on axigluon models can be
found in \cite{Gresham:2012kv} for ``light'' ($M_{G'}<450$ GeV)
axigluons and in \cite{Haisch:2011up} for heavy axigluons.  
The high-luminosity LHC should be able to rule out axigluon models currently under consideration,
though it is possible to come up with models that explain the Tevatron asymmetry and are difficult
to probe at the LHC.

\subsection{Other kinematic observables related to $A_{FB}$ at the LHC}
\label{toc:topkin-newObs}
The $A_{FB}$ asymmetry is only one of many angular variables whose
distributions can be measured in hadron collisions. 
Indeed,  if we consider $t \bar t$ production in parton collisions in semileptonic mode, 
the full kinematics of the event is characterized by 12 angles and the center-of-mass 
partonic collision energy.  In principle, kinematic distributions in these angles describe all
kinematic correlations in $t \bar t$ events and therefore are sensitive to potential deviations of
top couplings to $q \bar q$ or $gg$ initial states from their Standard Model values.  The 
forward-backward asymmetry provides an example of this more general framework. 

It will be worthwhile to pursue full angular analysis  of the $t \bar t$ events to understand subtle aspects of
top quark pair production or even processes with additional radiation, e.g.  $t \bar t j$,
especially in the context of studying top quark couplings to other Standard Model
particles, discussed in Section~\ref{sec:topcouplings}. Unfortunately, this general analysis has 
not been attempted so far. Here, we illustrate this general idea by mentioning
additional kinematic observables that can be explored. For example,
Refs.~\cite{Berge:2013xsa,Berge:2013csa} introduce two type of additional asymmetries in
$t \bar t j$ events that can be used to either probe the charge asymmetry or 
energy asymmetry in a complementary way or to 
provide additional tools to measure the $q g$ contribution to $t \bar t$ production.

\subsection{Kinematics at the linear collider}
\begin{figure}
\begin{center}
\includegraphics[width=0.47\textwidth]{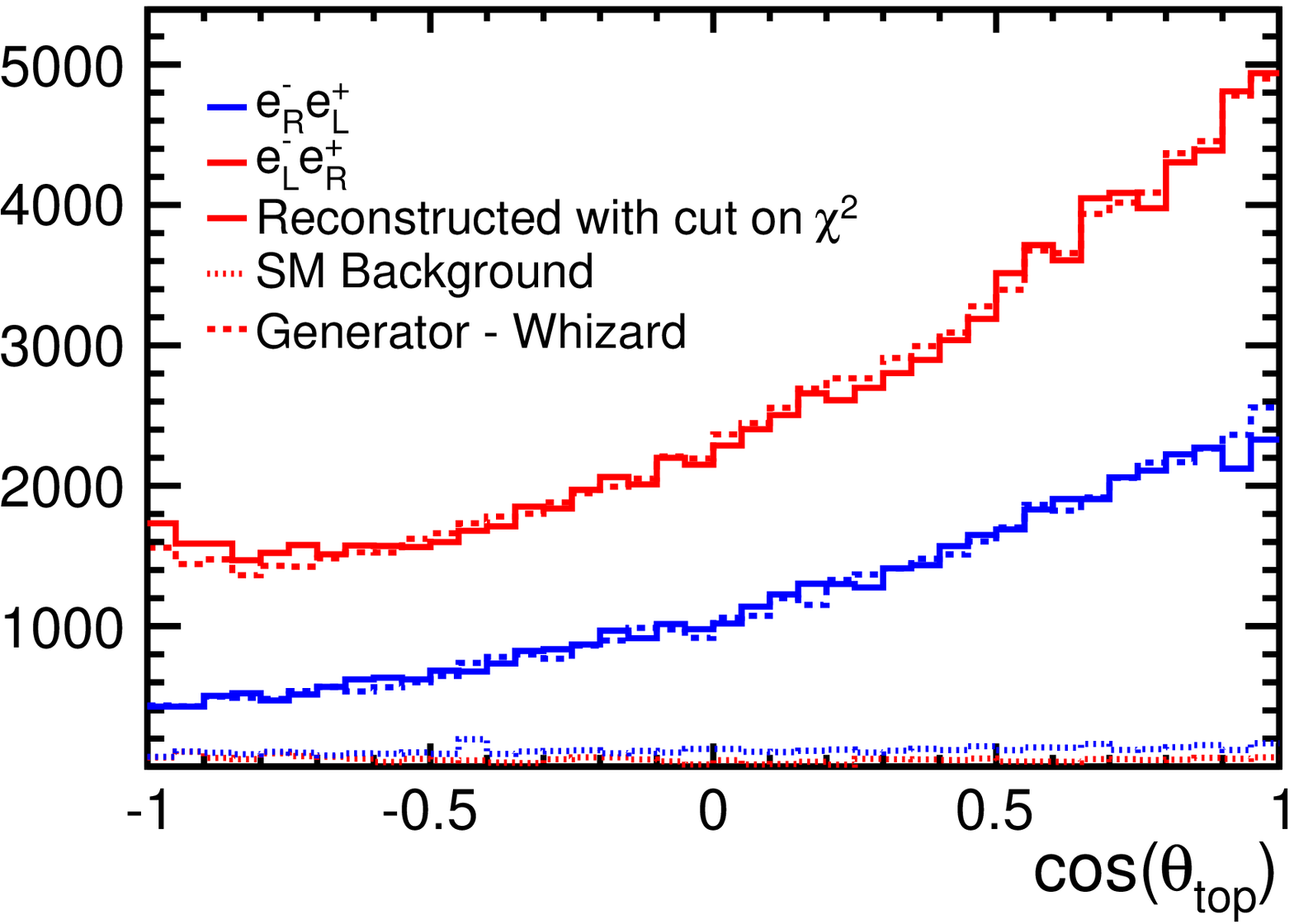}
\includegraphics[width=0.49\textwidth]{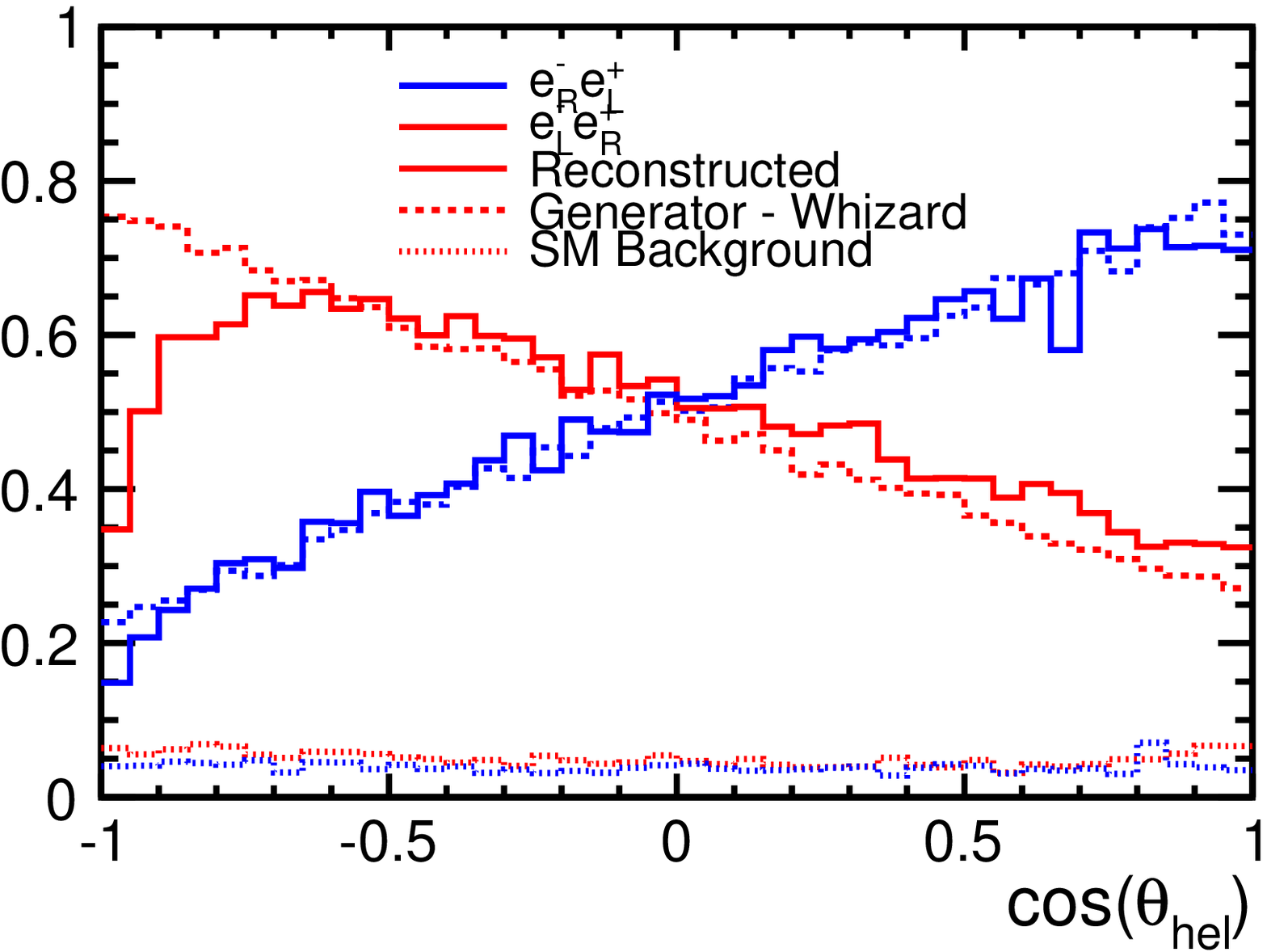}
\caption{\sl  \underline {Left:} Reconstructed forward backward asymmetry compared with the prediction by the event generator WHIZARD~\cite{Kilian:2007gr,Moretti:2001zz}. \underline{Right:} Polar angle of the decay lepton in the rest frame of the $t$~quark.}
\label{fig:afb_hel}
\end{center}
\end{figure}

At a linear collider, observables such as $A^t_{FB}$ or the slope of the helicity angle $\lambda_t$~\cite{Berger:2012nw} are sensitive to the chiral structure of the $t\overline{t} X$ vertex. A result of a full simulation study of semileptonic $t\overline{t}$ decays~\cite{bib:topcoupl-lalific} is shown in Fig.~\ref{fig:afb_hel}. 

The figure demonstrates that it will be possible to measure both the production angle $\theta_{top}$ of the $t$~quark and the helicity angle $\theta_{hel}$ to great precision over a large range, leading to measurements of $A^t_{FB}$ and $\lambda_t$ with a precision of about 2\%. Additionally, the $A^t_{FB}$ and other measurements of the $t\overline{t}$ system, will benefit from a $>60$\% pure sample \cite{Devetak:2010na} in which to measure the $b$ quark charge.  The chiral structures of couplings can be possibly be probed in this way.

Since a significant fraction of top studies will be around the $t \bar t$ threshold, understanding  kinematic distributions 
of top quark decay products in this region is important. This is a non-trivial 
problem that is affected by the need to account for 
QCD Coulomb interactions  to all orders.  While results for the total threshold 
cross-section $e^+e^- \to t \bar t$ are currently known through NNLO in QCD \cite{Hoang:2000yr}, 
similar  accuracy for kinematic distributions has not been achieved and it is an interesting and important problem 
to pursue in the future, if the potential of the threshold scan at the LHC is to be fully exploited.

\section{Rare decays} 
\label{sec:raredecays}

%
%

%
%

\subsection{Introduction}

%
%
%
Extensions of the SM often induce 
%
%
sizable
flavor-violating couplings between the top quark and other Standard Model particles, typically through new physics 
%
%
%
in loops. In contrast, flavor-changing neutral couplings of the top are highly suppressed in the SM, so that the measurement of anomalous or flavor-violating couplings of the top quark provides a sensitive probe of physics beyond the Standard Model. Since
the top quark decays before hadronizing, top flavor violation is ideally probed through direct flavor-changing neutral current (FCNC) production and decays of the top quark in experiments at the energy frontier.
Although flavor-violating couplings of the top may arise from many sources, if the responsible new physics is heavier than the top, it can be integrated out and its effects described by an effective Lagrangian: for details, see, for example, \cite{rarewp}.
In Section \ref{sec:theory} we summarize predictions for the size of flavor-changing top decays in the Standard Model and in various motivated models for new physics. In Section \ref{sec:limits} we collect the current best limits on top FCNC decays from direct searches.
%
%
In Section \ref{sec:future} we investigate the potential for future measurements at the LHC and ILC to constrain top FCNCs.

\subsection{Flavor-violating Top Decays} \label{sec:theory}

%
%
The
branching ratio (BR) of a flavor-violating decay of the top quark is
given
%
%
by the 
ratio
%
%
of the flavor-violating partial width relative to the dominant top quark partial width, $\Gamma(t \to bW)$. In Table \ref{tab:theory} we summarize predictions for top FCNC 
%
%
BRs 
in the Standard Model and various motivated new physics models. In the case of new physics models, the listed BR is intended as an approximate maximal value given ancillary direct and indirect constraints.

\begin{table}
\caption{SM and new physics model predictions for branching ratios of top FCNC decays. The SM predictions are taken from~\cite{AguilarSaavedra:2004wm}, on 2HDM with flavor violating Yukawa couplings~\cite{AguilarSaavedra:2004wm, Atwood:1996vj} (2HDM (FV) column), the 2HDM flavor conserving (FC) case from~\cite{Bejar:2006ww}, the MSSM with $1$TeV squarks and gluinos from~\cite{Cao:2007dk}, the MSSM for the R-parity violating case from~\cite{Yang:1997dk, Eilam:2001dh}, and warped extra dimensions (RS) from~\cite{Agashe:2006wa, Agashe:2009di} . } 
\begin{center}
\begin{tabular}{ccccccc}
\hline
Process & SM 
& 2HDM(FV) 
& 2HDM(FC) 
& MSSM 
& RPV 
& RS 
\\ \hline\hline 
$t \to Z u$ & $7 \times 10^{-17}$  & -- 			    	& -- 			& $\leq 10^{-7}$ & $\leq 10^{-6}$ & -- \\ 
$t \to Z c$ &  $1 \times 10^{-14}$  & $\leq 10^{-6}$	 & $\leq 10^{-10}$ & $\leq 10^{-7}$ & $\leq 10^{-6}$& $\leq 10^{-5}$ \\ 
$t \to  g u$ & $4 \times 10^{-14}$  & -- 				& -- & $\leq 10^{-7}$ & $\leq 10^{-6}$& -- \\
$t \to  g c$ &  $5 \times 10^{-12}$  & $\leq 10^{-4}$	 & $\leq 10^{-8}$ & $\leq 10^{-7}$& $\leq 10^{-6}$& $\leq 10^{-10}$\\
$t \to  \gamma u$ &  $4 \times 10^{-16}$& -- 			& -- & $\leq 10^{-8}$ & $\leq 10^{-9}$ & --\\
$t \to  \gamma c$ &  $5 \times 10^{-14}$ & $\leq 10^{-7}$ & $\leq 10^{-9}$ & $\leq 10^{-8}$ & $\leq 10^{-9}$ & $\leq 10^{-9}$\\
$t \to h u$ & $2 \times 10^{-17}$ & $6 \times 10^{-6}$ 	& -- & $\leq 10^{-5}$ & $\leq 10^{-9}$ & -- \\
$t \to h c$ & $3 \times 10^{-15}$ & $2 \times 10^{-3}$ 	& $\leq 10^{-5}$ & $\leq 10^{-5}$ & $\leq 10^{-9}$& $\leq 10^{-4}$ \\
   \hline\hline
\end{tabular}
\end{center}
\label{tab:theory}
\end{table}%

\subsubsection{SM top FCNCs}

SM contributions to top FCNCs are necessarily small, suppressed by both the GIM mechanism 
%
%
and by the large total width of the top quark due to the dominant mode $t \to bW$ \cite{Eilam:1990zc, Mele:1998ag}. This essentially guarantees that any measurable branching ratio for top FCNC decays is an indication of new physics. The values in Table \ref{tab:theory} are from the updated numerical evaluation in 
reference 
\cite{AguilarSaavedra:2004wm}. Note that the results are very sensitive to the value of $m_b$, since they scale as $m_b(m_t)^4$. The difference between decays involving $u$ quark and $c$ quarks arises from the relative factor $|V_{ub}/V_{cb}|^2$.

\subsubsection{BSM top FCNCs}

Many models for new physics predict new contributions to top FCNCs that are orders of magnitude in excess of SM expectations. Extended electroweak symmetry breaking sectors with two Higgs doublets (2HDM) lead to potentially measurable FCNCs. Parametric expectations are particularly large for 2HDM with tree-level flavor violation, for which flavor-violating couplings between Standard Model fermions and the heavy scalar Higgs $H$ or pseudoscalar $A$ 
are 
typically 
assumed 
%
%
to 
scale with quark masses, as
%
%
$\sqrt{m_q m_t / m_W^2}$, in order to remain consistent with limits on light quark FCNCs. The estimates in Table \ref{tab:theory} are taken from references \cite{Luke:1993cy, Atwood:1996vj}. The flavor-violating decays arise at one loop due to the exchange of $H, A$, and the charged Higgs scalar $H^\pm$, with the rate that depends on both the tree-level flavor-violating couplings between fermions and the heavy Higgs bosons and the masses of the heavy Higgs bosons themselves. 

Even when tree-level flavor conservation is guaranteed in the 2HDM by discrete symmetries, the model 
predicts measurable top FCNCs due to loop processes that involve the additional charged Higgs bosons. In this case the rate for flavor-violating processes depends on the mass of the charged Higgs and the angle $\tan \beta$ parameterizing the distribution of vacuum expectation values between the two Higgs doublets. In the Type-I 2HDM, the branching ratios are typically small; the most promising candidate is $t \to g c \sim 10^{-8}$, with rates for $t \to h q$ several orders of magnitude smaller. In the Type-II 2HDM, the leading contribution to $t \to h q$ is enhanced by $\mathcal{O}(\tan^4 \beta)$ and may be considerable at large $\tan \beta$. The most optimistic cases are presented in Table \ref{tab:theory}, taken from \cite{Bejar:2006ww} for Type I and Type II 2HDM. However, given that Higgs coupling measurements now constrain the allowed range of mixing angles in these 2HDM, the maximal rates for $t \to h q$ consistent with ancillary measurements are likely smaller. 

In the MSSM, top FCNCs arise at one loop in the presence of flavor-violating mixing in the soft SUSY-breaking mass matrices. Flavor violation involving the stops is much more weakly constrained by indirect measurements than flavor violation involving light squarks (particularly in the down-squark sector), allowing for potentially large mixing. However, rapidly-advancing limits on direct sparticle production have pushed the mass scale of squarks and gluinos to $\geq 1$ TeV, suppressing loop-induced branching ratios. To obtain realistic estimates, in Table \ref{tab:theory} we extrapolate the results of \cite{Cao:2007dk} to the case of $m_{\tilde g} \sim m_{\tilde q} = 1$ TeV. If $R$-parity is violated in the MSSM, top decays may also be induced at one loop by 
baryon 
($B$) 
or lepton
($L$)
number-violating RPV couplings. The effects of $B$-violating couplings are larger
%
%
by an order of magnitude or more. For the estimates in Table \ref{tab:theory}, we extrapolate the results of \cite{Yang:1997dk, Eilam:2001dh} to $m_{\tilde q} = 1$ TeV; for \cite{Yang:1997dk} we take their coupling parameter $\Lambda = 1$.

In models of warped extra dimensions,
%
%
top FCNCs arise when Standard Model fermions propagate in the extra dimension with profiles governed by the corresponding Yukawa couplings. These non-trivial profiles lead to flavor-violating couplings between SM fermions and the Kaluza-Klein (KK) excitations of the SM gauge bosons. Such couplings are largest for the top quark, whose profile typically has the most significant 
overlap with the gauge KK modes, and lead to flavor-violating couplings that depend on 5D Yukawa couplings and the mass scale of the gauge KK modes. Appreciable flavor-violating couplings involving the top quark and Higgs boson arise from analogous processes involving loops of fermion KK modes. 

{\bf A possible ``Discovery story''}: it is conceivable that the sensitivity of the LHC and the
ILC/CLIC to top FCNCs could lead to the discovery and identification of physics beyond the
Standard Model. An intriguing scenario is the observation of the flavor-violating decay $t \to Zc$ at the LHC with a branching ratio on the order of $10^{-5}$, at the limit of the projected high-luminosity reach. Such a branching ratio would be some nine orders of magnitude larger than the Standard Model expectation and a clear indication of new physics. At the LHC the primary backgrounds to this channel are Standard Model diboson $ZZ$ and $WZ$ production with additional jets, with a lesser component from $Z$+jets and rarer SM top processes $ttW$ and $ttZ$. The diboson backgrounds are fairly well understood and are in excellent agreement with simulations, and even such rare contributions as $ttW$ and $ttZ$ will be well-characterized by the end of the high-luminosity LHC run, making the observation of $t \to Zc$ fairly reliable.

A $t \to Zc$ signal described above is consistent with new physics arising from a variety of models, 
such as warped extra dimensions, a composite Higgs, or a flavor-violating two-Higgs-doublet model. Ancillary probes of FCNC processes become crucial for validating the signal and identifying its origin. Some of the most important probes that allow
differentiation between these options are the rare decays $t \to g c, t \to \gamma c$, and $t \to h c$, which have similar reach at the high-luminosity LHC. In the case of warped extra dimensions or a composite Higgs, the corresponding branching ratios for $t \to g c$ and  $t \to \gamma c$ are orders of magnitude below the sensitivity of the LHC, but the branching  $t \to h c$ may be as large as $10^{-4}$, within the reach of high-luminosity LHC. Thus a signal in $t \to Z c$ with a tentative signal in $t \to h c$ but no other channels would be indicative of warped extra dimensions or a pseudo-Goldstone composite Higgs, see Section~\ref{sec:newphysics}.
Such rates 
%
%
would also suggest a relatively low KK scale, so that complementary direct searches for heavy
resonances (see Section~\ref{sec:newphysics}) 
would play a crucial role in testing the consistency of this possibility. 
In addition, such a KK scale would also lead to up to a 
%
%
$10$ \% shift in $\bar{t} t Z$ coupling which can be probed at the
LHC or the ILC/CLIC (see section \ref{sec:topcouplings} of this report).
In contrast, in flavor-violating two-Higgs-doublet models, a visible $t \to Zc$ signal can be accompanied by comparable signals in $t \to g c$ and $t \to h c$, allowing this scenario to be similarly differentiated.

Complementary information can be provided by the ILC. Projections of the $\sqrt{s} = 500$ GeV ILC with 500 fb$^{-1}$ place its sensitivity to $t \to Z q$ coming from a $\gamma^\mu$ spin structure at the level of $10^{-4}$, but sensitivity to $t \to Z q$ in single top production from a $\sigma^{\mu \nu}$ structure at $\sim 10^{-5}$. The observation of comparable $t \to Zc$ signals at the LHC and ILC could then favor a $\sigma^{\mu \nu}$ coupling and rule out candidate explanations such as warped extra dimensions.

\subsection{Current Limits} \label{sec:limits}

Limits on various top FCNC decays have progressed rapidly in the LHC era. We summarize the current best limits from direct searches in Table \ref{tab:current}. CMS places the strongest limit on the decay $t \to Zq$ in the trilepton final state  
\cite{Chatrchyan:2012hqa} using the full 8 TeV data set. ATLAS sets a sub-leading limit on $t \to Zq$ using a portion of the 7 TeV data set, but also sets the leading limits on $t \to gq$ via a search for $s$-channel top production \cite{TheATLAScollaboration:2013vha} using 8 TeV data.  The Tevatron still maintains the best limits on some rare processes, in particular $t \to \gamma c$ from Run I  \cite{Abe:1997fz} and $t \to {\rm invisible}$ from Run II at CDF \cite{CDFinvis}. ZEUS maintains the best inferred limit on $t \to \gamma u$ \cite{Aaron:2009vv}. The Tevatron and HERA limits on $t \to \gamma q$ are expected to be superseded by LHC limits using the $7 \oplus 8$ TeV data set, but to date no official results are available. 

The recent discovery of the Higgs allows for limits to be set on $t \to hq$. The ATLAS collaboration sets the current best limit on $t \to hq$ using the diphoton decay of the Higgs with the full $7 \oplus 8$ TeV data set \cite{TheATLAScollaboration:2013nia}. In \cite{Craig:2012vj}  a limit was obtained on $t \to h q$ using the 7 TeV CMS multilepton search with 5 fb$^{-1}$ of data, assuming Standard Model branching ratios for a Higgs boson with $m_h = 125$ GeV. Similar limits may be set using the CMS same-sign dilepton search. The CMS multilepton search has recently been updated to $5 \oplus 20$ fb$^{-1}$ of $7 \oplus 8$ TeV data, and now includes $b$-tagged categories useful for constraining $t \to hq$; an official CMS search for $t \to hq$ using multi-leptons is ongoing.

\begin{table}
\caption{Current direct limits on top FCNCs. $(^*)$ denotes unofficial limits obtained from public results. The $q$ in the final state denotes sum over $q=u,c$.}
\begin{center}
\begin{tabular}{ccccc}
\hline\hline
Process & Br Limit & Search  & Dataset & Reference  \\ \hline
$t \to Z q$ & $7 \times 10^{-4}$ & CMS $t \bar t \to W b + Z q \to \ell \nu b + \ell \ell q$ & 19.5 fb$^{-1}$, 8 TeV & \cite{Chatrchyan:2012hqa} \\ 
$t \to Z q$ & $7.3 \times 10^{-3}$ & ATLAS $t \bar t \to W b + Z q \to \ell \nu b + \ell \ell q$ & 2.1 fb$^{-1}$, 7 TeV & \cite{Aad:2012ij} \\ 
$t \to  g u$ & $3.1 \times 10^{-5}$ & ATLAS $q g \to t \to Wb$ & 14.2 fb$^{-1}$, 8 TeV & \cite{TheATLAScollaboration:2013vha} \\
$t \to  g c$ & $1.6 \times 10^{-4}$ & ATLAS $q g \to t \to Wb$ & 14.2 fb$^{-1}$, 8 TeV &\cite{TheATLAScollaboration:2013vha}  \\
$t \to  \gamma u$ & $6.4 \times 10^{-3}$ & ZEUS $e^\pm p \to (t {\rm \; or \;} \bar t) + X$ & 474 pb$^{-1}$, 300 GeV & \cite{Aaron:2009vv}  \\
$t \to  \gamma q$ & $3.2 \times 10^{-2}$ & CDF $t \bar t \to Wb + \gamma q$ & 110 pb$^{-1}$, 1.8 TeV & \cite{Abe:1997fz}  \\
$t \to h q$ & $8.3 \times 10^{-3}$ & ATLAS $t \bar t \to Wb + h q \to  \ell \nu b +\gamma \gamma q$ & 20 fb$^{-1}$, 8 TeV &\cite{TheATLAScollaboration:2013nia}  \\
$t \to h q$ & $2.7 \times 10^{-2}$ & CMS$^*$ $t \bar t \to Wb + h q \to  \ell \nu b + \ell \ell q X$ & 5 fb$^{-1}$, 7 TeV &\cite{Craig:2012vj}  \\
$t \to$ invis. & $9 \times 10^{-2}$ & CDF $t \bar t \to Wb$ & 1.9 fb$^{-1}$, 1.96 TeV & \cite{CDFinvis} \\   \hline
\end{tabular}
\end{center}
\label{tab:current}
\end{table}%

Indirect limits on top FCNCs may also be set through single top production, $D^0$ oscillations, and neutron EDM limits. At present these limits are not competitive with direct searches at the LHC for final states involving 
photons and $Z$ bosons
%
%
\cite{Fox:2007in}, though they are comparable for final states involving $h$ \cite{Harnik:2012pb}.

%
%

%

\subsection{Projected Limits} \label{sec:future}

Although current direct limits on flavor-violating top couplings do not appreciably encroach on the parameter space of motivated theories 
(compare tables \ref{tab:theory} and \ref{tab:current}), 
future colliders should attain meaningful sensitivity 
%
%
(see table \ref{tab:LHCILC}). Here we will focus on the sensitivity of the $\sqrt{s} = 14$ TeV LHC after 300 and 3000 fb$^{-1}$ of integrated luminosity, as well as the ILC operating at $\sqrt{s} = 250$ and the ILC/CLIC at $500$ GeV, with 500 fb$^{-1}$ of integrated luminosity. The case of the $\sqrt{s} = 250$ GeV ILC is particularly interesting, since it possesses sensitivity to top FCNCs through single-top production via a photon or $Z$ boson.

\subsubsection{LHC projections}

At present, estimates of future LHC sensitivity to top FCNCs arise from two sources: official projections from the European Strategy Group (ESG) report 
\cite{ATLAS:2013hta} 
%
%
and approximate extrapolation from current searches at the 7 and 8 TeV LHC based on changes in luminosity, energy, and trigger thresholds. Table 
\ref{tab:LHCILC}
%
%
provides a summary of the projected limits at the 14 TeV LHC with 300 and 3000 fb$^{-1}$ integrated luminosity.

\begin{table}
\caption{Projected limits on top FCNCs at the LHC and ILC. ``Extrap.'' denotes estimates based on extrapolation as described in the text.
For the ILC/CLIC, limits for various tensor couplings (i.e., with $\sigma_{ \mu \nu }$ structure) are shown inside ().
}
\begin{center}
\begin{tabular}{ccccc}
\hline\hline
Process & Br Limit & Search  & Dataset & Reference  \\ \hline
$t \to Z q$ & $2.2\times 10^{-4}$ & ATLAS $t \bar t \to W b + Z q \to \ell \nu b + \ell \ell q$ & 300 fb$^{-1}$, 14 TeV &  \cite{ATLAS:2013hta} \\ 
%
%
$t \to Z q$ & $ 7 \times 10^{-5}$ & ATLAS $t \bar t \to W b + Z q \to \ell \nu b + \ell \ell q$ & 3000 fb$^{-1}$, 14 TeV &   \cite{ATLAS:2013hta} \\ 
%
%
$t \to Z q$ & $ 5 \, (2) \times 10^{-4 }$ & ILC single top, $\gamma_\mu$ ($\sigma_{\mu \nu}$) & 500 fb$^{-1}$, 250 GeV & Extrap.\\ 
 $t \to Z q$ & $ 1.5 \, (1.1) \times 10^{-4 \, (-5)}$ & ILC single top, $\gamma_\mu$ ($\sigma_{\mu \nu}$) & 500 fb$^{-1}$, 500 GeV & \cite{AguilarSaavedra:2001ab} \\ 
$t \to Z q$ & $ 1.6 \, (1.7) \times 10^{-3}$  & ILC $t \bar t$, $\gamma_\mu$ ($\sigma_{\mu \nu}$) & 500 fb$^{-1}$, 500 GeV & \cite{AguilarSaavedra:2001ab} \\ 
\hline
$t \to \gamma q$ & $8\times 10^{-5}$ & ATLAS $t \bar t \to W b + \gamma q$ & 300 fb$^{-1}$, 14 TeV & 
\cite{ATLAS:2013hta} \\ 
%
%
$t \to \gamma q$ & $ 2.5 \times 10^{-5}$ & ATLAS $t \bar t \to W b + \gamma q$  & 3000 fb$^{-1}$, 14 TeV &  
 \cite{ATLAS:2013hta} \\ 
%
%
$t \to \gamma q$ & $ 6 \times 10^{-5}$ & ILC single top & 500 fb$^{-1}$, 250 GeV & Extrap. \\ 
$t \to \gamma q$ & $ 6.4 \times 10^{-6}$ & ILC single top & 500 fb$^{-1}$, 500 GeV &\cite{AguilarSaavedra:2001ab} \\ 
$t \to \gamma q$ & $ 1.0 \times 10^{-4}$  & ILC $t \bar t$ & 500 fb$^{-1}$, 500 GeV & \cite{AguilarSaavedra:2001ab} \\ 
 \hline
$t \to  g u$ & $4 \times 10^{-6}$ & ATLAS $q g \to t \to Wb$ &300 fb$^{-1}$, 14 TeV  & Extrap. \\
$t \to  g u$ & $1 \times 10^{-6}$ & ATLAS $q g \to t \to Wb$ & 3000 fb$^{-1}$, 14 TeV  &  Extrap.  \\
$t \to  g c$ & $1 \times 10^{-5}$ & ATLAS $q g \to t \to Wb$ & 300 fb$^{-1}$, 14 TeV  &  Extrap. \\
$t \to  g c$ & $4 \times 10^{-6}$ & ATLAS $q g \to t \to Wb$ & 3000 fb$^{-1}$, 14 TeV  &  Extrap. \\
$t \to h q$ & $2 \times 10^{-3}$ & LHC $t \bar t \to Wb + h q \to  \ell \nu b + \ell \ell q X$ & 300 fb$^{-1}$, 14 TeV &  Extrap.\\
$t \to h q$ & $5 \times 10^{-4}$ & LHC $t \bar t \to Wb + h q \to  \ell \nu b + \ell \ell q X$ & 3000 fb$^{-1}$, 14 TeV &  Extrap. \\
$t \to h q$ & $5 \times 10^{-4}$ & LHC $t \bar t \to Wb + h q \to  \ell \nu b +\gamma \gamma q$ & 300 fb$^{-1}$, 14 TeV &  Extrap. \\
$t \to h q$ & $2 \times 10^{-4}$ & LHC $t \bar t \to Wb + h q \to  \ell \nu b +\gamma \gamma q$ & 3000 fb$^{-1}$, 14 TeV &  Extrap.  \\
 \hline

\end{tabular}
\end{center}
\label{tab:LHCILC}
\end{table}%

%
%
The ATLAS projections for $t \to q Z, \gamma$ are shown in the table. 
At present there is no public document from CMS with projections for 14 TeV sensitivity, nor are there official projections from either collaboration for $t \to gq$ or $t \to hq$. 

Estimates for LHC sensitivity to $t \to gq$ and $t \to hq$ are obtained by an approximate extrapolation from current searches accounting for changes in luminosity, energy, and trigger thresholds. While crude, when applied to $t \to Zq$ this procedure agrees to within $\mathcal{O}(10\%)$ with the official ATLAS ESG projections and so provides a useful benchmark in lieu of detailed study.
The $t \rightarrow h q$ entries in the table for the multilepton final state are derived from those in 
\cite{Craig:2012vj} by scaling with the luminosity and $t \bar t$ production cross-section.
This implies a 95\% CL limit ${\rm Br}(t \to h q) < 2 \times 10^{-3} (5 \times 10^{-4})$ with 300 (3000) fb$^{-1}$ at 14 TeV. 
Similarly, estimates based on \cite{TheATLAScollaboration:2013nia} give a sensitivity (95\% CL limit) in the $\ell \nu b + \gamma \gamma q$ final state of ${\rm Br}(t \to h q) < 5 \times 10^{-4} (2 \times 10^{-4})$ with 300 (3000) fb$^{-1}$ at 14 TeV. 
The extrapolation of $t \to gq$ is more delicate, since the process under study involves the $t gq$ anomalous coupling in the production mode. 
Using the results from \cite{Gao:2011fx} to extrapolate the observed 7 TeV limit \cite{Aad:2012gd} to 14 TeV, we find 
${\rm Br}(t \to g u) < 4 \times 10^{-6} (1 \times 10^{-6})$ with 300 (3000) fb$^{-1}$ at 14 TeV and
${\rm Br}(t \to g c) < 1 \times 10^{-5} (4 \times 10^{-6})$ with 300 (3000) fb$^{-1}$ at 14 TeV.

%
%
%

\subsubsection{Linear collider (ILC/CLIC) projections}

At the ILC/CLIC, sensitivity studies have focused on operation at $\sqrt{s} \geq 500$ GeV in order to probe both $e^+ e^- \to t \bar t, t \to Xq$ as well as the single top process $e^+ e^- \to t q$ due to, e.g., $t Zq$ or $t \gamma q$ anomalous vertices\footnote{As mentioned
in section \ref{sec:topcouplings}, TLEP has larger $t \bar{t}$ samples, but no polarization so that 
separating couplings to $\gamma$ from those to $Z$ will be difficult.}. Linear collider performance at $\sqrt{s} \geq 500$ GeV is studied in some detail in \cite{AguilarSaavedra:2001ab}, which forms the basis for sensitivity estimates quoted here. The study \cite{AguilarSaavedra:2001ab} includes 95\% CL estimates for various polarization options, including 80\% $e^-$ polarization and 45\% $e^+$ polarization, which are close to the polarization parameters 
planned
%
%
for the ILC.
%
%
In what follows we quote the 80\%/45\% polarization sensitivity, with the difference between 45\% $e^+$ polarization and 30\% $e^+$ polarization expected to lead to a small effect. We rescale the results of \cite{AguilarSaavedra:2001ab} to 500 fb$^{-1}$ to match the anticipated ILC/CLIC integrated luminosity; the results are presented in Table~\ref{tab:LHCILC}. Based on these estimates, ILC/CLIC sensitivity at $\sqrt{s} = 500$ GeV should be comparable to LHC sensitivity with 3 ab$^{-1}$ for $t \to Zq$ and $t \to \gamma q$. Since much of the sensitivity comes from single top production, the ILC/CLIC is less likely to provide comparable sensitivity to $t \to h q$ and $t \to g q$.

The ILC also provides sensitivity to $t Zq$ and $t \gamma q$ anomalous couplings at $\sqrt{s} = 250$ GeV through single top production via the $s$-channel exchange of a photon or $Z$ boson, $e^+ e^- \to t \bar c + \bar t c$. In fact, production via $Z$ exchange through the $\gamma_\mu$ vertex reaches its maximal cross-section
around 250 GeV and falls with increasing center-of-mass energy. Single top production cross-sections through $\gamma$ exchange or $Z$ exchange through the $\sigma_{\mu \nu}$ coupling grow with increasing energy but are still appreciable at $\sqrt{s} = 250$ GeV . The disadvantage of $\sqrt{s} = 250$ GeV relative to higher center-of-mass energies is primarily the larger SM backgrounds to the single-top final state. {\em In any event, this provides an intriguing opportunity for the ILC to probe new physics in the top sector even when operating below the $t \bar t$ threshold.}

The prospects for constraining $t Zq$ and $t \gamma q$ anomalous couplings at $\sqrt{s} = 250$ GeV have not been extensively studied, but we may extrapolate sensitivity reasonably well based on the results of \cite{Han:1998yr}. To obtain an estimate, we rescale the signal cross-section

 after cuts for  $e^+ e^- \to t \bar c + \bar t c$ via anomalous couplings  at $\sqrt{s} = 192$ GeV in \cite{Han:1998yr}  to $\sqrt{s} = 250$ GeV and conservatively assume the background cross-sections are similar between $\sqrt{s} = 192$ GeV and $\sqrt{s} = 250$ GeV. In actuality, the backgrounds should decrease with increasing center-of-mass energy. We assume a 60\% $b$-tag efficiency and arrive at 95\% CL estimates in Table~\ref{tab:LHCILC}.
%
%

%
%

\subsection{Vts and Vtd}
~
The measurement of the ratio of top decays with $b$-tagging to all top decays is sensitive to the
off-diagonal CKM matrix elements $V_{ts}$ and $V_{td}$~\cite{CMS-PAS-TOP-12-035}. A measurement of this
ratio at the sub-percent level should be possible at the high-luminosity LHC. The rapidity of
the top quark in $t$-channel single top quark production is also sensitive to $V_{ts}$ and 
$V_{td}$~\cite{AguilarSaavedra:2010wf}. The 
ultimate precision in $V_{ts}$ and $V_{td}$ will come from a combination of the 
two methods~\cite{Lacker:2012ek}.
%
%
Systematic uncertainties
and their correlations between different measurements will be a limiting factor, but a precision of 
better than 0.05 in $|V_{ts}|$ and $|V_{td}|$ should be achievable based on current studies.

\subsection{Summary}
%
%

Various well-motivated models predict branching ratios for top FCNC decays starting at $\sim 10^{-4}-10^{-5}$, with the most promising signals arising in two-Higgs-doublet models and various theories with warped extra dimensions. At present the LHC sensitivity to top FCNC decays is somewhat below the level predicted by motivated theories, with the notable exception of $t \to gu$ where searches for resonant single top production yield a limit $\mathcal{O}(10^{-4})$. 
However, future colliders, such as the 14 TeV LHC and $\sqrt{s} = 250$ ILC or 500 ILC/CLIC, provide meaningful sensitivity to flavor-violating couplings of the top quark, of the same order as the largest rates predicted in motivated theories. 
%
%
%
The LHC and the 
ILC/CLIC can be complementary in this regard: while the sensitivities in $t q Z/\gamma$ are (roughly) comparable
for the two colliders, the LHC is better for gluon couplings, but the ILC/CLIC is the way to go for 
probing the spin-structure of couplings.
Intriguingly, even at $\sqrt{s} = 250$ GeV the ILC should provide sensitivity to $t \to Zq, \gamma q$ that is
comparable to that
%
%
of the high-luminosity LHC.
Finally, going from LHC to HL-LHC can improve reach by roughly a factor of two (in rates).

%
%

\section{Searches for new particles associated with the top quark}
%
%
%
\label{sec:newphysics}

The top quark
%
%
provides a sensitive probe for physics beyond the Standard Model, based on the following
%
%
argument:
the presence of new physics at the TeV scale is very well-motivated by its role in solving the 
Planck-weak hierarchy problem of the SM. Namely, such 
new particles 
%
%
can prevent 
quantum corrections from dragging the Higgs boson mass (and hence its vev, i.e., the weak scale) all the way up to Planck scale.
Such new particles
must then necessarily couple to the Higgs boson.
However, 
because the top quark has the largest coupling (among SM particles) to the Higgs boson, quantum
%
%
corrections due to the top quark
are the  
%
%
dominant source of 
destabilization
%
%
of the weak scale. 
Thus,
such new particles typically also couple
preferentially to 
%
%
the top quark (among the other SM particles).

In this section, we focus on the {\em direct} production of such new particles, followed by their decay into top-like final states.
%
%
In most solutions to the Planck-weak problem, there are actually 
%
%
charge
$+2/3$, colored NP which accomplish the job of canceling the divergence from top quark loop
in the Higgs mass (and thus directly stabilizing 
%
%
the weak scale). These can be scalar/spin-0, i.e., 
stops in supersymmetry (SUSY: see review in \cite{Martin:1997ns}). 
%
%
Alternatively, they can be fermionic
(often denoted by ``top-partners''), as realized in little Higgs
(see reviews in \cite{Schmaltz:2005ky, Perelstein:2005ka}) and composite Higgs (accompanied by composite top) models. 
The latter are conjectured to be dual to the framework of a warped extra dimension, following the
AdS/CFT correspondence: see reviews in~\cite{Davoudiasl:2009cd, Contino:2010rs}).
%
%
The composite Higgs and top (or extra dimensional) models often also contain bosonic 
$t \bar{t}$ resonances.

With the above motivation, the studies performed for the Snowmass process can be grouped into the following three categories:
searches for stops, top-partners and $t \bar{t}$ resonances. These are described in turn below.
Note that virtual/indirect effects of such new particles also lead to rare/flavor changing neutral current decays of the 
top quark which are discussed in section \ref{sec:raredecays} of this report.
In addition, there can be shifts in couplings of the top quark which already exist in the SM (for example, flavor-preserving ones),  
as discussed in section \ref{sec:topcouplings} of this report.
Finally, these studies overlap with the work of the Snowmass Beyond Standard Model group, where further 
examples of New Physics related to top quarks can be found \cite{bsmreport}.

To illustrate an impact that top physics studies can have on discovering and understanding physics
beyond the Standard Model, we now describe a  {\bf discovery story}. Here the 
tell-tale signs for top and Higgs compositeness at the TeV scale is provided 
by a multitude of measurements and observations in top
physics~\cite{Davoudiasl:2009cd, Contino:2010rs}. 
In particular, suppose there is an observation of a shift in the $t \bar{t} Z$ cross-section of the order of
$10\%$ at a linear collider and, possibly even at the HL-LHC (see Section~\ref{sec:topcouplings}). It 
could be a ``smoking-gun'' signal for this scenario, since this model predicts such a size for this shift (as compared to
weakly coupled theories such as 
SUSY, where this effect is signficiantly suppressed and gives no signal at the LHC).
At the same time, in composite models rare decays $t \to c Z$ or $t \to c h$ can occur with a
branching fraction  of up to $10^{-4}$ which would then be accessible at the LHC and a lepton collider,
see Section~\ref{sec:raredecays}.
Finally, both top-partners and $t \bar{t}$ resonances with $O(\hbox{TeV})$ masses are
omnipresent in the compositeness scenario and would thus be accessible at the LHC, and especially
its high-luminosity option, via direct production.
Therefore, as our story illustrates, top quark physics at the LHC and at a linear collider may
be a crucial element of discovering and elucidating physics beyond the Standard Model.

\subsection{Stops}
\label{stops}

\def\t{\tilde{t}}
\def\n{\tilde{\chi}^0}
\def\go{\tilde{g}}
\def\c{\tilde{\chi}^\pm}

SUSY is perhaps the most popular solution to the Planck-weak hierarchy problem of the SM. It involves
addition of a {\em superpartner} for every particle of the SM, with a spin differing by $1/2$-unit from that of the corresponding
SM particle.
While, in general, superpartner masses in SUSY models are very model-dependent, naturalness strongly suggests that the scalar partners of the top quark, or {\it stops}, should have masses around the week scale.
The reason is that (as mentioned above) the stops 
cancel the largest divergence in the Higgs mass squared parameter, coming from the SM top loop.
This 
makes stops a prime target for LHC searches. 
The results of such searches are typically presented in terms of the ``vanilla stop'' simplified model, which contains two particles, a stop $\t$ and a neutralino LSP $\n$ (i.e., superpartner of photon and $Z$ or Higgs boson). The stop is assumed to decay via $\t\to t\n$ with a 100\% branching ratio. Within this model, the current ``generic'' bound on the stop mass is about 700 GeV~\cite{ATLAS8,CMS8}. 
One of the tasks of future experiments is to improve the reach on $m(\t)$ for generic spectra.
In fact, both ATLAS and CMS have presented estimates of the discovery reach of LHC-14 and HL-LHC in the vanilla stop model, extrapolating the present 1-lepton search~\cite{ATLAS:2013hta,CMS:2013xfa}. For a ``generic'' spectrum, stops up to approximately 800 (900) GeV can be discovered, at a 5-$\sigma$ level, with 300 fb$^{-1}$ (3 ab$^{-1}$) integrated luminosity.
It is interesting to determine if the reach at LHC 14 TeV for this generic case can be extended beyond
the above ATLAS/CMS projections using {\em special} techniques developed recently and so far applied only to the LHC 7/8 TeV.
The first study (as part of the Snowmass process) mentioned below is along these lines.

Moreover, it must be emphasized that lighter stops are still allowed by LHC 7/8 TeV. 
In particular: \\
\vspace{0.01in} \\
(a) If $m(\n)>250$ GeV, stops of any mass are allowed; \\
\vspace{0.01in} \\
(b) in the ``off-shell top'' region, $m_t > m(\tilde{t})-m(\n) > m_W$, stops above 300 GeV are allowed; \\
\vspace{0.01in} \\
(c) in the ``compressed'' region, $m(\tilde{t})\approx m(\n) +m_t$, stops of any mass are allowed (this includes the particularly challenging ``stealthy'' region,  $m(\tilde{t})\approx m(t) \gg m(\n)$); and \\
\vspace{0.01in} \\
(d) in the ``squeezed'' region, $m(\tilde{t})-m(\n) < m_W$, stops of any mass are allowed. \\

\noindent In all of these regions, the kinematics of stop production and decay yields events with little missing transverse energy (MET), reducing the efficiency of LHC searches. 
Thus, another goal of future experiments should be to explore the special regions listed above.
A couple of studies to cover the stealth 
stops of case (c) above were done 
as part of Snowmass process and are outlined below.

Although LHC will clearly play a leading role in the generic case\footnote{direct production of stops
at the ILC in this region is not possible, given the current bounds},
it should be emphasized that in any of the special regions, stops can still be within the kinematic reach of the ILC/CLIC, at $\sqrt{s}=500$ GeV or 1 TeV. In this case, the ILC could play a crucial role in discovering the stops and precisely determining/confirming their properties, {\it e.g.} spin and masses. 
%
%

Finally, {\em addition} of particles (such as gluino or chargino, i.e., superpartners of SM gluon or $W$) 
to the above simplified model is well-motivated. 
Adding such particles to the model generally weakens the exclusion limits.
Studies along these lines
were also performed for the Snowmass process and are described below.

\subsubsection{Vanilla stops}

Here fully hadronic decays using strategies inspired by~\cite{Plehn:2010st,Plehn:2012pr,Kaplan:2012gd,Dutta:2012kx}
are considered.
The fully hadronic channel has two advantages over leptonic searches. The first is that it has the largest branching fraction for the top decays. The second is that it has no inherent missing energy from neutrinos, so all the missing energy comes from the neutralinos. This allows many backgrounds to be reduced by vetoing events with leptons. 
Jet-substructure based 
top tagging (see section \ref{sec:detector} of this report) is used to distinguish signal from background.
The results are summarized in table~\ref{tab:vanilla}: for more details, see reference \cite{Stolarski:2013msa}.

\begin{table}[t]
\centering
\begin{tabular}{|c|c|c|c|c|}
\hline
	Collider	&	Energy		&	Luminosity		&		Cross Section &  Mass   \\       \hline		\hline
	LHC8 	&	8 TeV		&	20.5 fb$^{-1}$			&  10 fb     &	650 GeV                \\ \hline
	LHC 	&	14 TeV			&	300 fb$^{-1}$			&  3.5 fb     &	1 TeV                \\
	HL LHC	&	14 TeV		&	3 ab$^{-1}$			&  1.1 fb   &	1.2 TeV    	\\
	HE LHC &    	33 TeV                         &       3 ab$^{-1}$                &   91 ab   &      3.0 TeV     \\	
	VLHC &    	100 TeV                         &       1 ab$^{-1}$                     &    200 ab     &      5.7 TeV     \\	
\hline
\end{tabular} \hspace{-0.138cm}\vline
\vspace{0.3cm}
\caption{Stop production mass limits. The first line gives the current bound on stops.
The remaining lines give the estimated 5 $\sigma$ reach, based on a study for the Snowmass process, in stop pair production cross-section
and mass for different future hadron collider runs.  }
\label{tab:vanilla}
\end{table}%

\subsubsection{Stealth stops}
\label{sec:stops}

In the above-mentioned ATLAS/CMS projections of reach for stops at LHC 14 TeV, significant gaps in the coverage remain: for example, no discovery is possible for the LSP mass above 500 GeV, as well as in the compressed and stealthy regions, even at HL-LHC. It is clear that novel search strategies will be needed to cover these regions. 

Two studies of such strategies were contributed to our working group (see table \ref{tab:stealth} for summary of results). 
%
%
Reference
~\cite{Han:2013lna} focused on the stealthy stop region, which is particularly challenging since, unlike the region with a heavy neutralino, no significant MET is generated even in the presence of ISR jets. The challenge is to distinguish $\t\t^*$ events from a much larger $t\bar{t}$ background. Two methods to achieve this task have been studied: (a) using spin correlations, which are present in $t\bar{t}$ but not in $\t\t^*$ events, due to $\t$ being a scalar particle~\cite{Han:2012fw} (see also section \ref{sec:topkin-spincorrel} of this report); and (b) using an $m_{T2}$ cut in dileptonic event sample~\cite{Kilic:2012kw}. 
It was found that, using spin correlations, LHC-14 with 100 fb$^{-1}$ of data will be able to discover the stealthy 
stop at the $5\sigma$ level, assuming the stop mass of up to 200 GeV
and considering statistical errors only. 
Assuming a 15\% systematic error,
the $m_{T2}$ method will be able to discover right-handed stealthy stops {\em except} in the $(185, 195)$ GeV window. 
The sensitivity to the 
left-handed stop is poor, since there is no $m_{T2}$ tail in the signal in this case.

The second study \cite{wpS1} analyzed the possibility of using the vector boson fusion stop production channel, which provides additional jets that could be used to tag events with stealthy, compressed, or light stops. For compressed stops with $\Delta M = m(\tilde{t}) - m(\tilde{\chi}^0) = 10$ GeV, it was found that the LHC-14 with 300 fb$^{-1}$ of data will be able to probe $m(\tilde{t}) = 400$ GeV at a $5 \sigma$ level. The mass reach increases for larger $\Delta M$. Studies for stops in the stealthy, ``off-shell top'', and ``squeezed'' regions using vector boson fusion are ongoing.

%
%
%

\begin{table}[t]
\centering
\begin{tabular}{|c|c|c|c|}
\hline
	Collider		&	Luminosity		&  Technique &   Mass reach
	%
	%
	\\       \hline		\hline
	LHC	14 TeV 		& 100 fb$^{-1}$			&   spin-correlations    &	175--200 GeV 
	%
	%
	 (for stealth stops, \\
  & & & statistical uncertainty only)               \\ \hline
	LHC	14 TeV 		& 100 fb$^{-1}$			&   dileptonic $m_{ T2 }$   & 
175--185~GeV and 195-200~GeV 
	%
	%
	(for stealth tops)          \\ \hline
	LHC 	14 TeV			& 300 fb$^{-1}$			&   VBF   &	400 GeV  
	%
	%
	(for compressed case, \\
  & & &  with $m(\tilde{t}) - m(\tilde{\chi}^0) = 10$ GeV)             \\ \hline
\end{tabular} \hspace{-0.138cm}
\vspace{0.3cm}
\caption{Estimated 5 $\sigma$ discovery reach for stealth/compressed stops, based on various studies for the Snowmass process. See the text for explanation of these concepts.}
\label{tab:stealth}
\end{table}%

\subsubsection{Gluino-initiated stop production}

In addition to stops, naturalness also strongly motivates a light gluino, constraining its mass through the one-loop QCD correction to stop mass. A rough naturalness bound is $m(\go)<2m(\t)$~\cite{Brust:2011tb}. This motivates considering a simplified model with gluino, stop and an LSP, with a decay $\go\to t\bar{t}+$MET. Assuming that this decay proceeds via an off-shell stop and has a 100\% branching ratio, LHC-8 searches rule out gluino masses up to about $1.3$ TeV, provided that the LSP mass is below 500 GeV~\cite{ATLAS8go,CMS8go}. Extrapolating the search in the all-hadronic channel, CMS estimates a $5\sigma$ discovery reach of $1.7$ TeV at LHC-14 with 300 fb$^{-1}$ of data~\cite{CMS:2013xfa}. 
%
%
%
For gluino masses above TeV, boosted tops become increasingly common in $\tilde{g}$ decays, and can be used to tag SUSY events~\cite{Berger:2011af} (see section \ref{sec:detector} of this report for techniques to identify boosted tops). 
A preliminary study (with no detector simulation) 
%
%
in reference~\cite{wpmaxim1} suggests that a search using top tags, in combination with more traditional kinematic cuts in all-hadronic channel, at the LHC-14 with 300 fb$^{-1}$ (3 ab$^{-1}$) of data will be able to discover gluinos up to $1.8$ ($2.1$) TeV, provided that the stop mass is below $1.1$ ($1.4$) TeV.

\begin{table}[t]
\centering
\begin{tabular}{|c|c|c|}
\hline
	Collider		&	Luminosity		&   Mass
	%
	%
	\\       \hline		\hline
%
%
LHC 	14 TeV 			&	300 fb$^{-1}$			    &	 1.8 TeV                \\ \hline
	LHC 	14 TeV 			&	3000 fb$^{-1}$			    &	 2.1 TeV                \\ \hline
\end{tabular} \hspace{-0.138cm}
\vspace{0.3cm}
\caption{Estimated 5 $\sigma$ discovery reach for gluino decaying into stops, with $R$-parity conservation, based on a study for the Snowmass process.}
\label{tab:}
\end{table}%

\subsubsection{Including more electroweak particles}

Another well-motivated extension of the vanilla stop simplified model is to add a chargino $\c$, with $m(\c)<m(\t)$. This is also motivated by naturalness, since the charged Higgsino mass is controlled by the $\mu$ parameter which cannot be far above 100 GeV in natural SUSY 
models \cite{Brust:2011tb}. This simplified model has the possibility of {\it asymmetric} stop events: {\it e.g.} $pp\to \t\t^*, \t\to t\n, \t^* \to b\c$. A study of the LHC sensitivity to this signal was performed: for details, see reference \cite{wpG}. The proposed search uses the 1-lepton+MET channel, and relies crucially on the ``topness'' variable, introduced in~\cite{Graesser:2012qy} as a general tool to suppress the $t\bar{t}$ background in this channel. 
It was found that $5\sigma$ discovery is possible at LHC-14 with $300$ fb$^{-1}$ for stop masses up to about 1 TeV, if $m(\n)$ is below about 400 GeV. With $3000$ fb$^{-1}$, the discovery reach extends to stop masses about 1.3 TeV for similar $\n$. 

A related simplified model was used in the study in reference \cite{wpS2}. Motivated by the ``well-tempered neutralino'' dark matter scenario~\cite{ArkaniHamed:2006mb}, this study considered a spectrum with light bino and Higgsino, leading to three neutralino and one chargino states at the bottom of the SUSY spectrum. It was assumed that all these states are lighter than the stop. The analysis focused on the dilepton signature, where the leptons can come either from top decays or from $\chi^0_{2,3}\to Z\chi^0_1$. It was found that the reach is about 700 GeV. 

\begin{table}[t]
\centering
\begin{tabular}{|c|c|c|c|}
\hline
	Collider		&	Luminosity		&  Technique/channel &     Mass
	%
	%
	 \\    \hline		\hline
	LHC	14 TeV 		& 300 fb$^{-1}$			&     topness, asymmetric &	1 TeV                \\ \hline
	LHC 	14 TeV 				&	3000 fb$^{-1}$			&    topness, asymmetric  &	1.3 TeV                \\ \hline
	LHC 	14 TeV 			&	300 fb$^{-1}$			&    dilepton, well-tempered neutralino  &	
	700 GeV$^{\rm preliminary}$                \\ \hline
\end{tabular} \hspace{-0.138cm}
\vspace{0.3cm}
\caption{Estimated 5 $\sigma$ discovery reach for stops decaying into chargino, based on various studies for the Snowmass process. See the text for explanation of these concepts.}
\label{tab:chargino}
\end{table}%

\subsubsection{$R$-parity violation}

Yet another interesting scenario is $R$-parity violating (RPV) supersymmetry, where 
decay modes are modified relative to the above cases of $R$-parity conservation.
For example, a stop can decay via $\t\to \bar{b}\bar{s}$ induced by the $UDD$ superpotential operator. This scenario emerges naturally from models with minimal flavor violation~\cite{Nikolidakis:2007fc,Csaki:2011ge}. 
Direct stop production in this case yields all-hadronic final states, 
%
%
but it might still be possible to search in this channel: see, for example, the Snowmass study \cite{Duggan:2013yna}.
However, just as in conventional SUSY, naturalness strongly suggests the presence of relatively light gluinos. Gluino decays via cascades involving stops, $\go\to \t t, \t \to 2j$, may be observable, even though they do not produce large MET. If $\go$ is Majorana, as in simplest SUSY models, such decays can provide a striking same-sign dilepton (SSDL) signature. Current SSDL searches already rule out gluinos up to 800 GeV, independent of the stop mass, in the RPV scenario~\cite{Berger:2013sir}. 
%
%
At LHC-14 with 300 fb$^{-1}$ (3 ab$^{-1}$) of data, the projected 5 $\sigma$ reach of this search in gluino mass is $1.4$ ($1.6-1.75$) TeV, depending on the stop mass~\cite{Saelim:2013gea}.
These estimates include fast detector simulation with Delphes~\cite{Delphes3}, but no pile-up.
%
%
It also 
includes an improvement in sensitivity due to an additional requirement of one or two massive jets. The massive jets can be either due to boosted stop decays, or to accidental mergers of neighboring jets in a high jet multiplicity signal event.
An alternative approach is a search in a single-lepton channel, which has a higher rate and applies to both Majorana and Dirac gluinos~\cite{Han:2012cu}. In this case, the requirement of stop mass reconstruction from jet pairs can be used as an additional handle to suppress backgrounds. At the 14~TeV LHC, this search will be sensitive to {\em Dirac} gluino masses up to about $2$ TeV~\cite{wpK}.

\begin{table}[t]
\centering
\begin{tabular}{|c|c|c|c|}
\hline
	Collider		&	Luminosity		&  Technique &   Reach  \\       \hline		\hline
	LHC	14 TeV 			& 300 fb$^{-1}$			&    same-sign dilepton   & 1.4 TeV                \\ \hline
	LHC	14 TeV 			& 3000 fb$^{-1}$			&    same-sign dilepton   & 1.6--1.75 TeV                \\ \hline
	LHC 	14 TeV			&	300 fb$^{-1}$			&  single-lepton, reconstruct mass   &	2 TeV $^{\rm preliminary}$  
	\\
	            \hline
\end{tabular} \hspace{-0.138cm}
\vspace{0.3cm}
\caption{Estimated 5 $\sigma$ reach for gluino decaying into stops, with $R$-parity violation, based on various studies for the Snowmass process. See the text for explanation of these concepts.
%
%
}
\label{tab:RPV}
\end{table}%

\subsection{Top-partners}

As mentioned above, in non-SUSY solutions to the Planck-weak hierarchy problem,
the divergence in Higgs mass squared parameter from SM top loop is canceled
by new {\em fermions} which are vector-like under the SM gauge symmetries. Typically, they are color triplets with electric charge 2/3 (i.e., same as the SM top and hence these new particles 
are dubbed top-partners.
Such 
particles can also arise in other extensions of the SM, so it is useful to follow a model-independent, simplified approach
in studying their signals.
The top-partners can be produced via QCD interactions in pairs or singly \cite{Mrazek:2009yu}, 
the latter resulting from coupling of
top-partner to SM top/bottom, as needed to cancel the SM top divergence in Higgs mass squared parameter.
%

%
Based on the $SU(2)_L$ gauge symmetry of the SM, the top-partners are often accompanied by ``bottom-partners''.
Finally, in some composite Higgs models, an extension of the EW symmetry group (from that in the SM) is motivated by
the goal of avoiding constrains from $Z b \bar{b}$ \cite{Agashe:2006at}: this results in the appearance of color triplet, but charge $5/3$ particles
(in addition to the above top/bottom partners).

%
%

In short, there are three types of vector-like quarks which are well-motivated extensions
of the SM, namely, top and bottom-partners and charge-$5/3$ fermions.
Once produced, these vector-like quarks can decay into a 
top-like final state.
All of these cases were studied for various LHC scenarios as part of the Snowmass process (including
both single and pair production of top-partners mentioned above) and are discussed below.
Note that the current (LHC 7/8 TeV) bound on these quarks is at least 800 GeV \cite{CMS-PAS-B2G-12-015,
ATLAS-CONF-2013-060} so that 
their direct production at the ILC is not possible.

The single production mechanisms have larger rates than those of pair productions for 
{\em heavier} top/bottom partners.  Moreover, analyses of single-production channels 
might permit the measurement of the 
%
%
the 
couplings of top-partners to the SM top/bottom quark (which were mentioned towards the end of the first paragraph of this section).
Note that the  
%
%
top-partner single-production, that proceeds via the intermediate exchange of a bottom quark, 
has a rate significantly higher than those for a single bottom partner or a charge-$5/3$ 
%
%
partner, which are mediated by the exchange of a top. Hence, in the following, for bottom partners 
and charge $5/3$ quarks, only pair production is considered.
All samples used for the top-partner studies were processed with the {\sc Delphes}~\cite{Delphes3} fast
detector simulation, using the generic Snowmass detector parameters~\cite{SnowmassDetSim}. The
background samples were generated in bins of $H_T$, as described in~\cite{SnowmassBackgrounds, SnowmassOSG}. 

{\bf Pair production of top-partners:} 
The top-partner has three possible decay modes: 
$bW$, $ t H$ and $ Z t $.  The interesting feature is that, in the limit of a heavy top-partner, 
the decay modes are equally shared by these three modes 
(following the principle of Goldstone equivalence). A general analysis 
such that the three branching fractions $bW$, $ t H$, and $ Z t $ are 
free parameters, subject to the constraint that they add up to unity
and span a ``triangle'' of branching fractions, has been performed recently~\cite{CMS-PAS-B2G-12-015}.
However, it is useful to consider a nominal set of branching fractions, 
one that is motivated by naturalness with BF($T\rightarrow Wb$) = 0.5, 
BF($T\rightarrow th$) = BF($T\rightarrow Zt$) = 0.25.
The results from an analysis~\cite{CMS-PAS-B2G-12-015} based on lepton+jets and multi-lepton signatures for all 
decay modes $bW$, $ t H$ and $ Z t $, and utilizing presence of multiple $b$-jets, large number of jets, 
are given in  table \ref{tab:topbPartner}. 
Due to the large mass of the top-partner,
jet substructure techniques are utilized to identify the $W$-tagged and top-tagged jets and to keep the 
signal yield high and fully  optimize the signal-to-background discrimination. 
Reference \cite{Bhattacharya:2013iea} contains more details of the analysis.


{\bf Single production of top/bottom-partners:} 
As mentioned above, the top-partner 
%
%
can decay into one of three possible final states: $h t$,  $Zt$ and $W b$. Since the $W +$ jet backgrounds are considerable for the third mode, here the focus is on the first two decay modes.
The basic idea is to reconstruct the top-partner mass using the full event information in the decays
$ht\rightarrow b\overline{b}\ell\nu b$ and $Zt\rightarrow \ell\ell\ell\nu b$.
The results are summarized in Table~\ref{tab:topbPartner},
for more details, see reference \cite{Andeen:2013zca}.


{\bf Pair production of bottom-partners:} 
The decays of bottom-partners 
can be into $W^- t$, $Zb$, or $Hb$. 
Thus, pair production of bottom partners can lead to interesting signal of 
same-sign dileptons via $W^- t W^+ \bar{t} \rightarrow b \bar{b} \; 2 \; W^+ \; 2 W^-$, 
followed by leptonic decays of both $W^+$ (or $W^-$).
More details of this study can be found in reference \cite{Varnes:2013pxa}; here, 
only the final results, obtained using the nominal  branching fractions 
BF$(B \rightarrow Wt)$ = 0.5, BF$(B \rightarrow Zb)$ = BF$(B \rightarrow Hb)$= 0.25, are shown in Table~\ref{tab:topbPartner}.


{\bf Pair production of Charge-$5/3$ fermion:} 
The charge-$5/3$ vector-like fermions 
($T_{5/3}$)~\cite{Contino:2008hi} decay via $T_{5/3} \rightarrow t W^+ \rightarrow b W^+ W^+$
and thus the pair production of these particles results in same-sign dileptons 
with a branching fraction of approximately  0.2. These final states can be distinguished from 
Standard Model backgrounds with same-sign dileptons by the presence of 
jets or leptons from the second $T_{5/3}$ and by the magnitude of the 
scalar sum of transverse momenta of the decay products.
At $T_{5/3}$ masses of interest, hadronically $W$ bosons and top quarks from the $T_{5/3}$ decay 
are often highly boosted and can  be identified using the tagging methods described in 
Section~\ref{sec:boostedtop} which enhance the background 
discrimination. Furthermore, it is possible to fully reconstruct the second $T_{5/3}$, 
in case of fully hadronic decay, and compute its mass. 
Table \ref{tab:topbPartner} displays the reach for
these exotic quarks; for more details, see reference\cite{Avetisyan:2013rca}.

%

\begin{table}[t]
\centering
\begin{tabular}{|c|c|c|c|c|c|}
\hline
Collider      &	Luminosity   & Pileup & 3$\sigma$ evidence & 5$\sigma$ discovery  & 95\% CL \\ \hline		\hline
\multicolumn{6}{|l|}{\bf top-partner pair production} \\
LHC 14 TeV   &	300 fb$^{-1}$	&    50  &   1340~GeV & 1200~GeV&	1450~GeV     \\ \hline
LHC 14 TeV   &	3 ab$^{-1}$	&   140  &   1580~GeV & 1450~GeV&	1740~GeV \\ \hline
LHC 33 TeV   &  3 ab$^{-1}$     &    140  &   2750~GeV & 2400~GeV  & 3200~GeV\\	\hline \hline
\multicolumn{6}{|l|}{\bf top-partner single production} \\
LHC 14 TeV    &	300 fb$^{-1}$	&    50  & 1275~GeV & 1150~GeV   & \\ \hline
LHC 14 TeV    &	3 ab$^{-1}$	&   140  & 1130~GeV & 1000~GeV	    &     \\ \hline
LHC 33 TeV    &	3 ab$^{-1}$	&   140   & 1350~GeV & 1220~GeV  &	\\ \hline
LHC 100 TeV   &       3 ab$^{-1}$ &  50  &   1750~GeV & 1600~GeV  &	        \\	\hline
LHC 100 TeV   &       3 ab$^{-1}$ &  140  &   1750~GeV & 1575~GeV	 & \\	\hline\hline
\multicolumn{6}{|l|}{\bf bottom-partner pair production}\\
LHC 14 TeV	&	300 fb$^{-1}$   &    50 &	1210~GeV   & 1080~GeV & 1330~GeV  \\ \hline
LHC 14 TeV	&	3 ab$^{-1}$     &   140 &	1490~GeV & 1330~GeV & $>$1500~GeV  \\ \hline
LHC 33 TeV      &       300 fb$^{-1}$   &  50 &   $>$ 1500~GeV & $>$ 1500~GeV & $>$ 1500~GeV \\	\hline\hline
\multicolumn{6}{|l|}{\bf Charge $5/3$ fermion pair production}\\
LHC 14 TeV    &    300 fb$^{-1}$ &    50  & 1.51 TeV           & 1.39 TeV           & 1.57 TeV \\ \hline
LHC 14 TeV    &    3 ab$^{-1}$   &   140  & 1.66 TeV           & 1.55 TeV           & 1.76 TeV \\ \hline 
LHC 33 TeV    & 3 ab$^{-1}$   &   140  & 2.50 TeV           & 2.35 TeV           & 2.69 TeV \\ \hline
\end{tabular}
\vspace{0.3cm}
\caption{Expected mass sensitivity for heavy top and bottom partners, based on various studies for the Snowmass process.}
\label{tab:topbPartner}
\end{table}%

\subsection{$t \bar{t}$ resonances}

As mentioned earlier, in 
%
%
non-supersymmetric solutions to the Planck-weak
hierarchy problem, there are typically 
bosonic new particles which decay dominantly into $t \bar{t}$.
Examples are leptophobic $Z^\prime$'s in topcolor models ~\cite{Harris:1999ya} 
or KK gluons in warped extra dimensional frameworks (conjectured to be dual 4D
composite Higgs models:  see reviews in \cite{Davoudiasl:2009cd, Contino:2010rs}).
Moreover, such $t \bar{t}$ resonances are favored to be rather
heavy (a few TeV) due to the constraints from various precision tests.
and/or by the current direct bounds from LHC 7/8 TeV \cite{Aad:2013nca, CMSttbar1, Aad:2012raa, CMSttbar2}.
Thus, the 
top quarks resulting from their decays are boosted so that the top decay products
can be quite collimated, requiring special identification techniques which have
been developed recently (for more details, see section \ref{sec:detector} of this report). 
%
%
In some models, these $t \bar{t}$ resonances can also be broad, thereby adding to the challenge of 
searching for them.

Three such studies of discovery of $t \bar{t}$ resonances were done as part of the Snowmass process and are discussed in what follows.
Of course, post-discovery, the focus will shift to determination of the quantum numbers
of these $t \bar{t}$ resonances.
For example, the spin and chiral structure of couplings of these resonances 
can be measured via angular distribution and polarization of the resulting top quarks: see, for example, references \cite{Agashe:2006hk, Barger:2006hm, Baumgart:2011wk}.
Finally, note that given the mass range of these $t \bar{t}$ resonances, ILC/CLIC would not play a role
in a direct search.

\subsubsection{Dileptonic}

One of the studies (reference \cite{Iashvili:2013ada}) focused on $W$'s from both top quarks decaying into lepton (called ``dileptonic'' $t \bar{t}$). 
One expects hadronic activity near the leptons, since the boost of the top quark can put a lepton and a $b$-jet into the same cone. 
So, SM $t \bar{t}$ background can be suppressed by in fact  requiring {\em smaller} separation between lepton and the 
closest jet.
The results are summarized in Table \ref{tab:dileptonic}.

\begin{table}[t]
\centering
\begin{tabular}{|c|c|c|c|c|}
\hline
	Collider		&	Luminosity		&		Pileup &  95 \% exclusion & 5 $\sigma$ discovery  \\       \hline		\hline

		LHC 14 TeV			&	300 fb$^{-1}$			&   50  &	4.4 TeV &   2.8 TeV             \\ \hline
		LHC 14 TeV		&	3 ab$^{-1}$			&   140 &	4.7  TeV    & 4.1 TeV	\\ \hline
%
\end{tabular} \hspace{-0.138cm} 
\vspace{0.3cm}
\caption{Expected mass sensitivity for a $Z^{ \prime }$ decaying into {\em di}leptonic $t \bar{t}$, based on a study for the Snowmass
process.}
\label{tab:dileptonic}
\end{table}%

\subsubsection{Semileptonic and fully hadronic}

Alternatively, one of the two $W$'s from the decays of the top quarks can give a lepton, while
the other one decays into hadrons (semileptonic $t \bar{t}$) or none of the two $W$'s 
decays into leptons (fully hadronic $t \bar{t}$).
A study for the Snowmass process of the fully hadronic channel 
%
%
utilized $b$-tagging and large-$R$ jet substructure to distinguish jets from boosted top quarks from jets from QCD
multijet production, and evaluated the prospects for a search for narrow resonances.
The results are expressed in terms of both a leptophobic $Z^{ \prime }$ and KK graviton in warped extra dimensional models: see 
tables \ref{tab:yale1} and for details, see sections 3.2 and 3.3.2 of reference \cite{Agashe:2013fda}.
The study for Snowmass process of the semileptonic channel is still to be completed and so the results
shown in  \ref{tab:yale2} (for leptophobic $Z^{ \prime }$ and KK gluon) are from the ATLAS Snowmass 
whitepaper~\cite{ATLAS:2013hta}.

\begin{table}[t]
\centering
\begin{tabular}{|c|c|c|c|c|}
\hline
	Collider		&	Luminosity		&		Pileup &  95 \% exclusion for $Z^{ \prime }$  &  95 \% exclusion for KK graviton\footnote{The numbers shown are for right-handed top quark being localized near TeV brane and with 
	the parameter $k / M_{ \rm Pl } = 1$.}
	\\       \hline		\hline

		LHC 14 TeV			&	300 fb$^{-1}$			&    50  &	3.7 TeV & 2 TeV           \\ \hline
		LHC 14 TeV		&	3 ab$^{-1}$			&   140 &	4.1 TeV &  2.8 TeV  	\\ \hline
\end{tabular} \hspace{-0.138cm} 
\vspace{0.3cm}
\caption{Expected mass sensitivity for a leptophbic $Z^{ \prime }$ and KK graviton
decaying into 
%
%
fully hadronic $t \bar{t}$, based on a study for the Snowmass process. 
%
}
\label{tab:yale1}
\end{table}%

\begin{table}[t]
\centering
\begin{tabular}{|c|c|c|c|c|}
\hline
	Collider		&	Luminosity		&		Pileup &  95 \% exclusion for $Z^{ \prime }$  
	& 95 \% exclusion for KK gluon
	%
%
%
	\\       \hline		\hline

		LHC 14 TeV			&	300 fb$^{-1}$			&    50  &	 3.3 TeV 
		& 4.3 TeV           
		\\ \hline
		LHC 14 TeV		&	3 ab$^{-1}$			&   140 &	5.5  TeV 
		&   6.7 TeV 
		 	\\ \hline
\end{tabular} \hspace{-0.138cm} 
\vspace{0.3cm}
\caption{Expected mass sensitivity for a leptophobic $Z^{ \prime }$ 
and KK  {\em gluon}
decaying into 
{\em semi}leptonic 
%
%
$t \bar{t}$~\cite{ATLAS:2013hta}.
%
%
}
\label{tab:yale2}
\end{table}%

Another study focused on the KK gluon (which is typically a {\em broad} resonance) in warped extra dimensional models.
In order to identify boosted top quarks, it used 
the Template Overlap Method (TOM) \cite{Almeida:2010pa}.
TOM has been extensively studied in the past in the context of theoretical studies of boosted tops and boosted Higgs 
decays \cite{Backovic:2012jj}, as well as used by the ATLAS collaboration for a boosted resonance search \cite{Aad:2012raa}. The method is designed to match the energy distribution of a boosted jet to the parton-level configuration of a boosted top decay, with all kinematic constraints taken into account. 
%
%
Low susceptibility to intermediate levels of pileup (i.e. 20 interactions per bunch crossing), makes TOM particularly attractive for boosted top analyses at the LHC.
For more details about how the TOM is used in this study, see section 3.4 of reference \cite{Agashe:2013fda}: the results are
shown in Table \ref{tab:TOM}.

\begin{table}[t]
\centering
\begin{tabular}{|c|c|c|c|c|}
\hline
	Collider		&	Luminosity		&		Pileup &  3 $\sigma$ evidence & 
	%
	%
	 5 $\sigma$ discovery \\       \hline		\hline

		LHC 14 TeV			&	300 fb$^{-1}$			&    50  &	3.8 TeV & 3.2 TeV         \\ \hline
		LHC 14 TeV		&	3 ab$^{-1}$			&   50 &	4.4 TeV & 3.5  TeV  	\\ \hline
%
%
\end{tabular} \hspace{-0.138cm} 
\vspace{0.3cm}
\caption{Expected mass sensitivity for a KK gluon decaying into semileptonic
%
%
$t \bar{t}$, based on a study for the Snowmass process using
the template overlap method.}
\label{tab:TOM}
\end{table}%

\subsubsection{Single-top resonance}
~
Resonances can appear not only in top pair production, but also in single top quark production. This
final state is particularly sensitive to a high-mass $W'$~boson that couples primarily to 
quarks. Current limits for $W'$ production are around
1.8~TeV~\cite{Aad:2012ej,ATLAS-CONF-2013-050,Chatrchyan:2012gqa}. A Snowmass study shows that the reach
for $W'$ can be extended to 5~TeV (6~TeV) with 300~fb$^{-1}$ (3000~fb$^{-1}$) at the 14~TeV 
LHC~\cite{WKKgWhitePaper}.

In warped extra dimensional models, the KK gluon discussed in the previous section can also have a
sub-dominant decay into $ t \bar{c}$ (and $\bar{t} c$)~\cite{Aquino:2006vp}. This process is also
relevant for the flavor sector, see the chapter on Flavor working group report~\cite{flavorWG}.
The final state has a single top quark, just like $W'\rightarrow tb$, but now the other quark jet
is from a charm quark rather than a bottom quark. This has consequences for the $b$-tag multiplicity
and background suppression. The Snowmass study finds a mass limit on the KK gluon of about 3.5~TeV if the
branching ratio to $tc$ is 20\%. If this branching ratio is less than 5\%, the signal is buried
below backgrounds and no limit can be set.

A fourth-generation quark with chromomagnetic couplings will be visible in the single top plus $W$~boson
final state~\cite{Nutter:2012an,Aad:2013rna}. Due to the strong nature of the 
%
%
fourth-generation bottom quark 
production process,
the reach for this particle at the high-luminosity LHC should be multi-TeV, similar to the $W'$.

\section{Top Algorithms and Detectors}
\label{sec:detector}
Studies of top quarks at future colliders will, in many cases, require dealing  with new
environments.  These include the increased number of pile-up events per bunch crossing in
the high-luminosity phase of the LHC, and an increasing reliance on boosted techniques for top
identification as the higher energy of the LHC and stronger constraints on scale of BSM physics 
 will require exploration of  higher
invariant mass events in top quark pair production. In this section we discuss how existing
algorithms for top quark studies fare in these cases, and whether or not physics studies that
we described in the preceding sections are in fact viable given difficult experimental environments of 
future colliders.  We find that the largest gains from high-luminosity running will be made
by taking advantage of boosted top identification algorithms, and by using jet grooming techniques, 
both of which look promising even in
high pile-up environments.  We also discuss the unique experimental conditions  
of the linear collider for top quark studies.

\subsection{Top quark identification at low transverse momentum}
\label{sec:unboosted}
~
Many of the top quarks produced at the LHC have low transverse momenta, where
$p_\perp$ is in the range $25-50$~GeV.  Measurements of 
the total and differential $t\bar{t}$ cross-sections (Sections~\ref{sec:kinematics}
and~\ref{sec:topcouplings}), of the top-quark mass (Section~\ref{sec:topmass}),
charge asymmetry (Section~\ref{sec:kinematics}), and single-top measurements
(Section~\ref{sec:topcouplings})
all require precise and efficient reconstruction of top quarks at low transverse momenta.
Top-quark reconstruction at low transverse energies is limited by a
number of factors that determine total systematic uncertainty, including:
a) jet-energy scale uncertainty which typically  accounts for
$50\%$ of the overall uncertainty in  traditional top-quark measurements based on jets; 
b) jet-energy resolution uncertainty;
c) $b$-tagging efficiency uncertainty and mistag rates; and 
d) uncertainty on missing transverse-energy reconstruction.
This indicates that progress in precision top measurements that involve jet
reconstruction at low $p_\perp$ will require a better 
understanding of low-$p_{T}$ jets and $b-$tagging.

The high-luminosity upgrade of the LHC will have an important impact on low-$p_\perp$ top
physics. In the current design, we expect more than $100$ pileup events per bunch crossing, which
will have a negative impact on many final-state observables, particularly on low-$p_T$ jets and
$b$-tagged jets due to their large associated systematics.
Studies of this scenario  \cite{TopAlgWhitePaper}  were performed 
for $pp$ collisions at a center-of-mass energy of $\sqrt{s}=14$~TeV using 
a fast detector simulation
based on the {\sc Delphes} 3.08 framework~\cite{Ovyn:2009tx}.
Jets are reconstructed at the LHC using the anti-$k_T$ algorithm~\cite{Cacciari:2008gp}
with distance parameter $R=0.4$ (ATLAS) and $R=0.5$ (CMS and Snowmass-specific studies).
These high-luminosity MC simulation studies showed that, in general, pileup events
deposit energy in many calorimeter cells and hence shift the raw jet transverse
energies by approximately $50~(120)$~GeV for $50~(140)$ pileup events,
adding about one additional GeV for each pileup event. This energy needs to be subtracted
jet-by-jet using average energies deposited elsewhere in the calorimeter. Tracking
in jet reconstruction is also useful, not only to refine the jet energy measurement but
also to mitigate the impact of pileup.
Nevertheless, the subtraction of pileup results in smeared jet transverse momenta.
In addition, there will be many 
low-$p_T$ fake jets created from pileup events. While tracking can be used to address
some of these issues as well, pileup also creates many additional tracks that need to be separated
from the tracks belonging to each jet in an event.

\begin{figure}
\begin{center}
\subfigure[]{
\includegraphics[scale=0.35, angle=0]{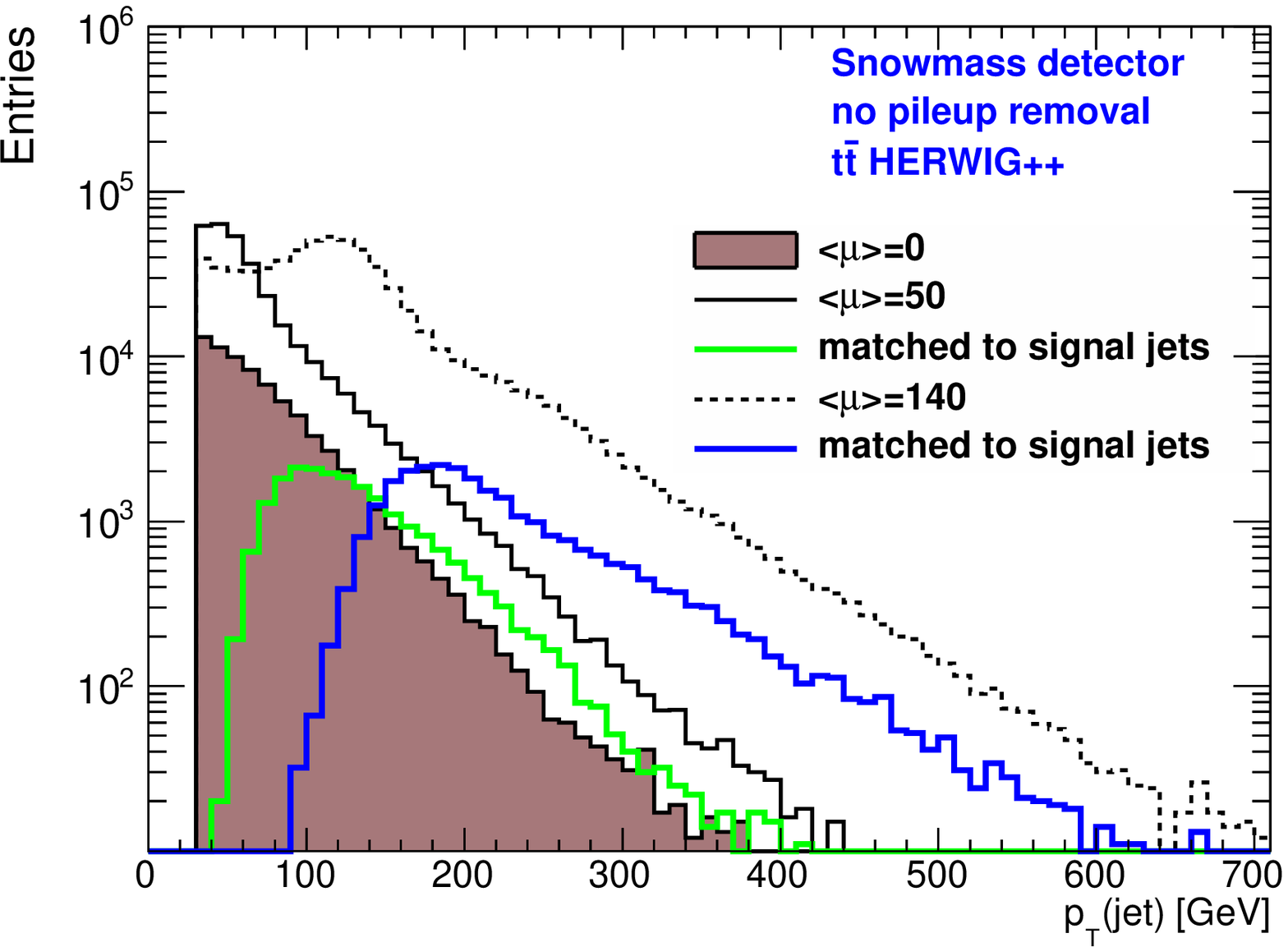}
}
\subfigure[]{
 \includegraphics[scale=0.33, angle=0]{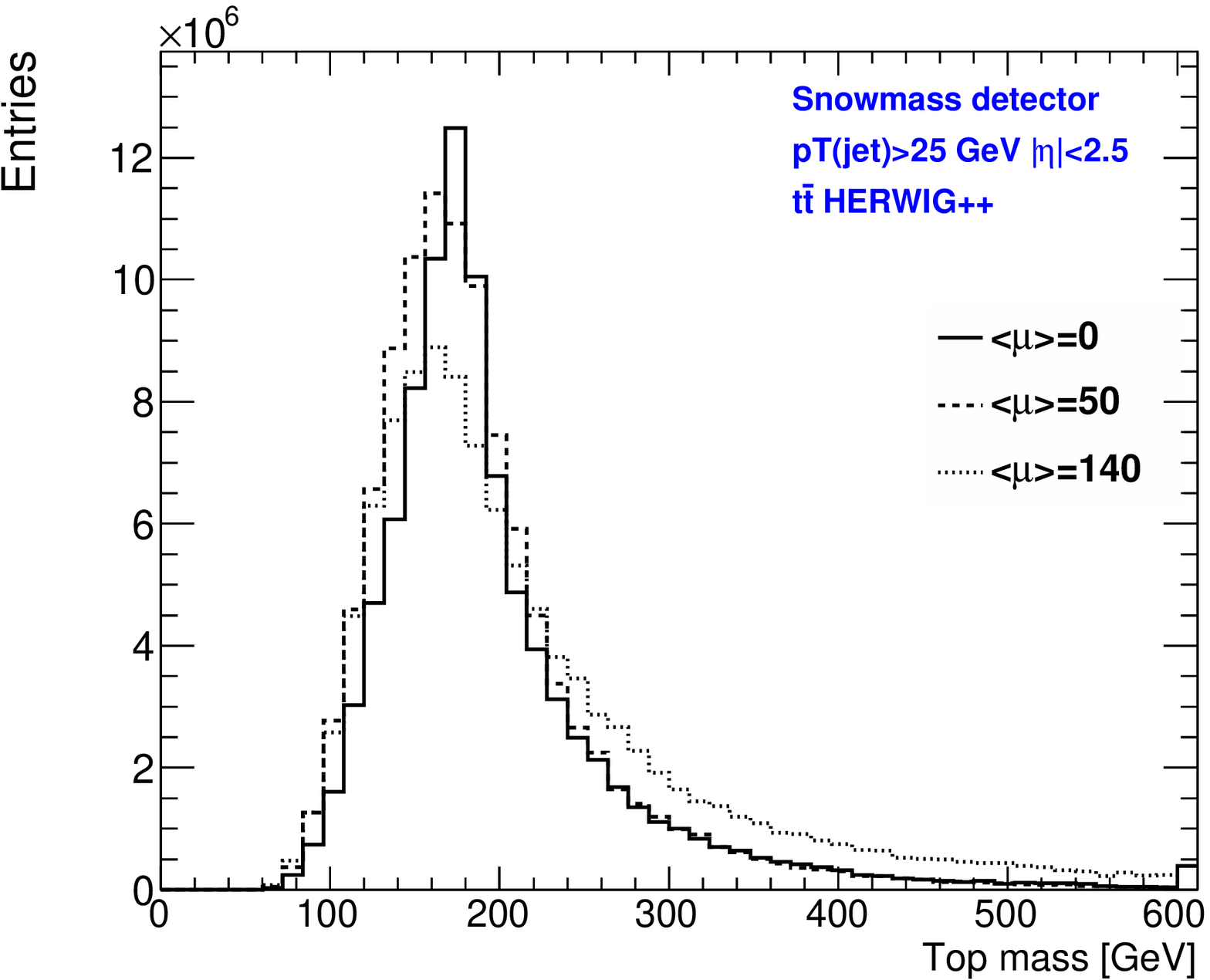}
}
\caption{(a) Plots of jet $p_T$ distributions for different pileup scenarios using the {\sc Delphes} simulation.  Also 
shown are only the jets matched to the top quarks in the event for each pileup scenario, demonstrating 
the large effect of additional pileup events on top quark reconstruction. (b) Reconstructed top quark masses from trijets by requiring at least four jets with $p_T >25$~GeV and $|\eta|<$ 2.5,  and  at least one of the jets must be tagged by a $b$-tagging algorithm.}
\label{pileupjets}
\end{center}
\end{figure}

Figure~\ref{pileupjets}(a) shows the effect of different pileup scenarios on the jet $p_T$
distribution.
One consequence of the energy shift is that for the selection of top quark signal jets,
pileup subtraction techniques will likely correct energies of the signal jets by 200-400\%, leading 
to larger uncertainties compared to previous analyses.  
Uncertainties due to pileup will become dominant, and are expected to increase by a factor 
of two or more at the highest LHC luminosity. 
As an example, a $2\%$ jet-energy scale uncertainty 
for a jet measurement without pileup 
translates to a $3\%$($5\%)$ uncertainty in case of 50 (140) pileup events scenario.  

Since uncertainties in jet resolution, jet energy scales,
and $b$-tagging are dominant uncertainties in many measurements related to top quarks,
it is to be expected that precision of such measurements will not improve at higher
luminosities and will deteriorate unless new jet energy calibration methods are adopted.
While data-driven techniques are likely to improve the assessment of the jet energy scale,
it is unlikely that this can make a significant difference to the above conclusion. As
the result, the standard top mass measurements  do not improve at the high-luminosity 
phase of the LHC.  It should be noted, however, 
that new techniques which are less reliant on precise knowledge of jet energies 
will be able to take advantage
of the high statistics of the high-luminosity LHC, as we discuss in Section~\ref{sec:topmass}.

The reconstruction of the top quark mass that is used in many other top quark analyses
will also be degraded by the high pileup in high-luminosity runs. 
A {\sc Delphes} MC study shows that using the trijet mass for top-reconstruction is strongly
affected by pileup events even when particle-flow methods and pile-up subtraction
techniques are used to mitigate the problem~\cite{TopAlgWhitePaper}. Figure~\ref{pileupjets}(b) shows
the reconstructed top mass using a procedure similar to the one discussed
in~\cite{ATLAS-CONF-2011-120}.
It was also observed~\cite{TopAlgWhitePaper} that 
the trijet mass for top-reconstruction strongly depends on  top transverse momentum $p_{T}$
due to large jet multiplicity from  ISR/FSR. 
For $p_{T}>$700~GeV, the peak position is at 400~GeV, assuming the same transverse 
momentum cuts as for low-$p_T$ measurements. This may limit our ability to identify top quarks
at such large $p_{T}$ using the traditional low-energy approaches.

Runs at high pileup will also affect other top physics measurements, 
such as $t(\bar{t})$+jets and associated top production (such as  $Ht\bar{t}$),
discussed in Section~\ref{sec:topcouplings}, 
as well as  searches for new physics that require a good understanding of 
low-$p_{T}$ top quarks, for example searches for rare top decays 
(Section~\ref{sec:raredecays}).
Indeed, low-$p_T$ top quarks require the reconstruction of jets
with transverse momentum $30-100$~GeV, which are exactly the jets that are difficult to correct for pileup
effects.  However, for rare decays or other counting measurements, it may not be
so necessary to determine jet energy scales as precisely as measurements using kinematic
shapes, so the advantage of the large statistics may very well dominate.  Nevertheless,
these measurements will be affected by the reduced performance of $b$-tagging
at high pileup, so work will be needed in this area compared to existing algorithms.

In conclusion, we find that 
the high-luminosity $pp$ collision runs at 14~TeV with more than hundred pile-up events 
are unfavorable for high-precision top quark measurements based on jets with
transverse momenta below 100~GeV.
This conclusion will affect $t\bar{t}$ measurements with top quarks produced 
near threshold that rely on precise
knowledge of jet energies, but will affect rate-dependent measurements to a lesser extent, 
especially with improvements in $b$-tagging algorithms.
We believe that a combination of multiple measurements by CMS and ATLAS may lead to a
reduction of systematic uncertainties even in the high pile-up environment.

As will be discussed  below, the  high-luminosity LHC runs will affect studies of
high-$p_{T}(\mathrm{top})$ top quarks to a lesser extent.
It is therefore important  to discuss the future of  boosted top measurements, where additional  reconstruction
techniques can be utilized.

\subsection{Methods particular to boosted top quarks}
\label{sec:boostedtop}
~
As we explained in Section~\ref{sec:newphysics},  
top quarks play a very important role in
many searches for new particles at the highest energies.
We find that current algorithms for top quark identification at high-$p_T$ can lead to 
performance that is similar to what is achieved 
in current experiments, provided that some modifications to the reconstruction methods are implemented  or
detectors upgrades are performed.

\begin{figure}[tb]
\begin{center}

\subfigure[Anti-kt R=0.5]{
 \includegraphics[scale=0.32, angle=0]{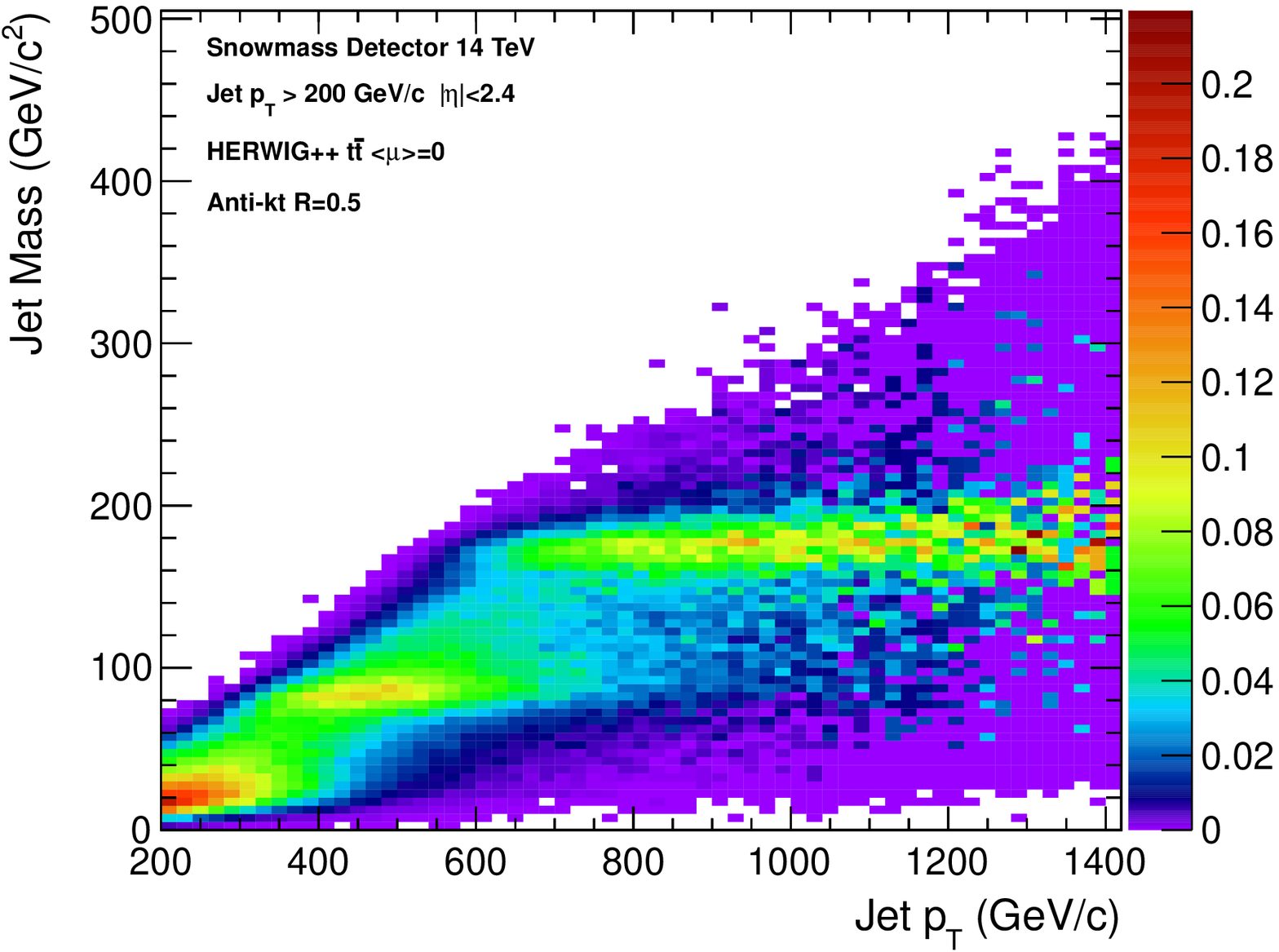}
 }
 \subfigure[Cambridge Aachen R=0.8]{
 \includegraphics[scale=0.32, angle=0]{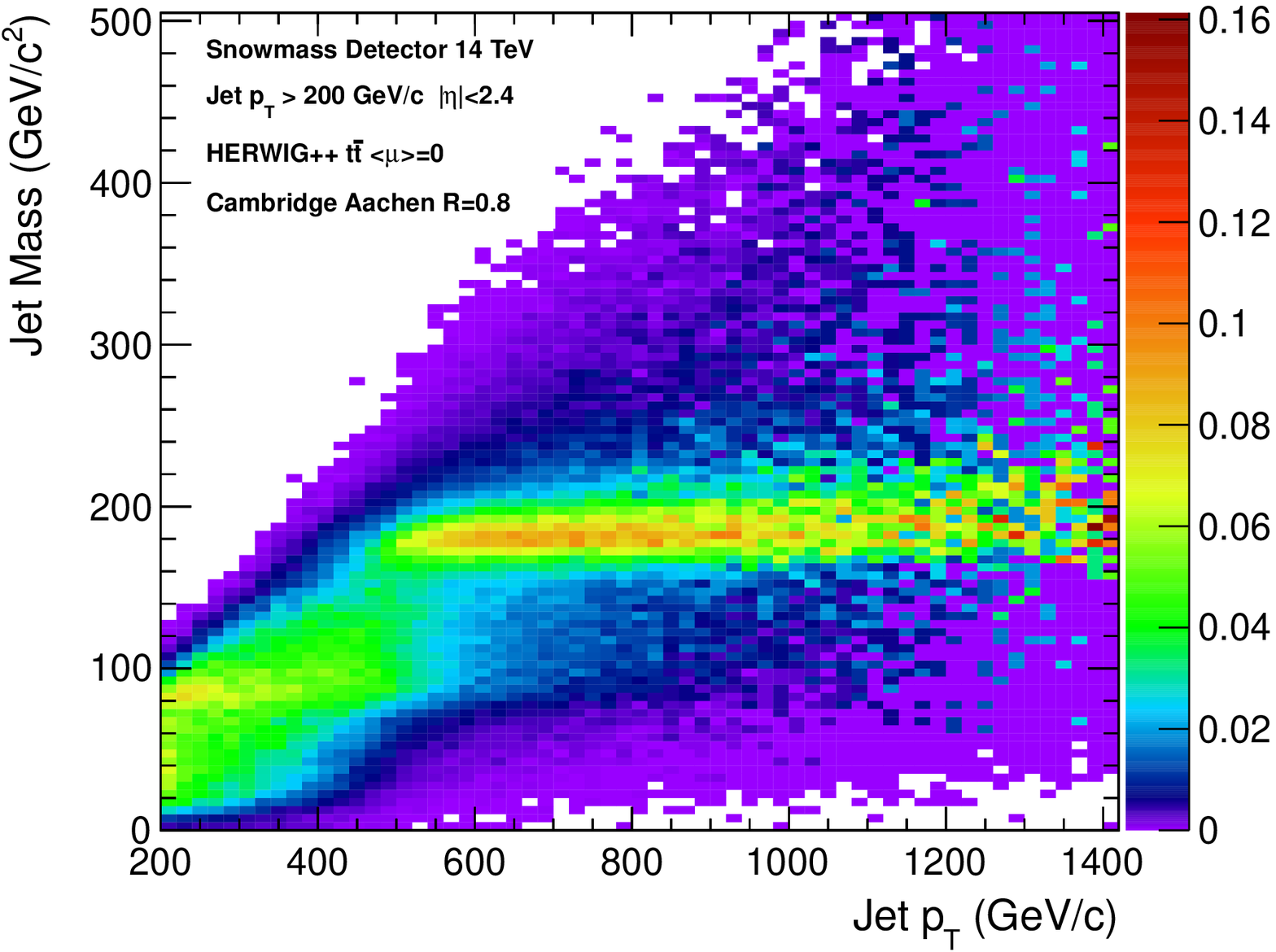}
 }
\end{center}
\caption{
Jet mass vs jet transverse momenta in the {\sc DELPHES} fast simulation for pp collisions at 
$14$ TeV for different jet algorithms. The jet transverse spectrum has been reweighed to be flat.
}
\label{fig:jetmass_pileups}
\end{figure}

The decay products of a top quark with high  $p_T$ are sufficiently collimated
to be reconstructed within a single jet.  This happens above $\sim 400$ GeV for jets with
$R=0.8$. Figure~\ref{fig:jetmass_pileups} shows the evolution of jet mass with the jet
transverse momentum for the $t\bar{t}$ process. Because all of the top decay products
fall within a single jet, specialized techniques involving jet substructure are 
required~\cite{Abdesselam:2010pt,Altheimer:2012mn}.
Semileptonic top decay reconstructions must introduce modified isolation criteria when the
lepton starts  to overlap with the $b$~quark jet from the top decay.
This reconstruction of the top mass within a single jet itself is a good discriminant
between boosted top quarks and the overwhelming background from QCD jet production. For
example, a recent study~\cite{Auerbach:2013by} has shown that a signal of boosted hadronic
top quarks from a $Z'$~boson decay can be observed in the jet mass distribution alone for jets
with $p_T > 800$~GeV. Discrimination can possibly be improved further with the addition of
$b$-tagging.
The reconstruction of the top jet through its proximity to the mass  of the top is the  
basic idea behind the boosted top studies.  In addition, further signal/background 
separation  is achieved by using specialized algorithms 
that split the top jet into sub-jets, and then examine those to determine if observed jet substructure 
is consistent with soft and collinear QCD radiation or with the decay of a heavy object into jets through a 
point-like interaction vertex.


{\bf Jet grooming. } Boosted jets are affected by pileup just like the unboosted ones 
discussed in Section~\ref{sec:unboosted}. Several algorithms, collectively known as jet
grooming algorithms, attempt to mitigate the effect of pileup on jet observables, such as jet
mass, by removing soft and wide-angle constituents of jets.  
The effect of three different jet grooming algorithms have been studied: 
pruning~\cite{Ellis:2009me, Ellis:2009su}, trimming~\cite{Krohn:2009th}, and 
filtering~\cite{butterworth-2008-100}.  The application of these jet grooming algorithms results
in a jet mass distribution that is relatively stable as the number of pileup events
increases. Additionally, the jet grooming procedures significantly reduce the
masses of QCD jets, enhancing signal/background  discrimination significantly.


{\bf Substructure and jet shapes.}
Jet substructure and jet shapes are often discussed as a useful tool for the identification
of top quarks and for reduction of the overwhelming rate from conventional QCD processes
\cite{Agashe:2006hk,Lillie:2007yh,Butterworth:2007ke,Almeida:2008tp,Almeida:2008yp,
Kaplan:2008ie,Brooijmans:2008,Butterworth:2009qa,Ellis:2009su,ATL-PHYS-PUB-2009-081,CMS-PAS-JME-09-001,Almeida:2010pa,Hackstein:2010wk,Chekanov:2010vc,Chekanov:2010gv,ATL-PHYS-PUB-2010-008}.
For example, the $N$-subjettiness algorithm~\cite{Nsubj} aims to determine the consistency of a
jet with a hypothesized number of subjets.
Such tools can give good discrimination between top quark jets and QCD jets, however,  such
discrimination degrades somewhat with the additional pileup activity.

It is also beneficial to identify the two subjets corresponding to the $W$~boson produced in
the top quark decay. Using trimming, a $W$ mass peak can be extracted which is relatively
stable even with 140 additional pileup events added. 


{\bf Top tagging.} In addition to the substructure quantities described above, there are
several algorithms (top taggers) which combine multiple jet observables to identify top jets
and provide additional discrimination from QCD jets.
Two top-tagging algorithms which are currently in use by experimental efforts include the CMS
Top Tagger~\cite{CMS-PAS-JME-09-001,Kaplan:2008ie} and the HEP Top 
Tagger~\cite{Plehn:2010st,Aad:2013gja,ATLAS:2012am,Aad:2012dpa,Aad:2012raa}.
The CMS top tagger decomposes a jet into up to 4 subjets.
Then requirements on the jet mass ($140 < m_j < 250$ GeV), number of subjets (3 or more) and 
a quantity which is a proxy for the mass of the $W$~boson within the jet (minimum pairwise
subjet mass $>50$ GeV) , are imposed to isolate boosted top quarks.  
We have studied the effect of pileup on the efficiency of the CMS top tagger.  
With no additional pileup events, the efficiency of the algorithm maintains its maximum value
of $\sim 40$\% up to jet $p_T$ values of 1.2-1.3 TeV, at which point the efficiency begins to
fall to 10\% or lower for jets with $p_T > 1.5$ TeV.  This efficiency drop at high $p_T$ can
be ameliorated by increasing calorimeter granularity, but extra radiation off of the top quark
also affects the algorithm in the very high $p_T$ regime.  With additional pileup events (and no
correction applied to the subjets), the efficiency degradation happens at much lower $p_T$
values.
The rate of QCD jets passing the algorithm is also affected.  With no additional pileup events,
this mistag rate remains below 5\% over the entire range of jet $p_T$.  After adding 140 
pileup interactions to the simulated events, the mistag rate from QCD jets increases to a
maximum of 45\% at a $p_T$ of 500~GeV, though this can be reduced through additional algorithm
improvements.  However, above approximately $p_T > 1$ TeV we expect that there will be minimal
impact after pile-up corrections.  


{\bf Detector effects.} At large values of the top quark $p_T$, such as the region above
1.5~TeV at the LHC, QCD radiation as well as the size of the detector elements become
a limiting factor.  In this regime, top quarks will have hard radiations that may be
identified as subjets and the top quark decay products become so highly collimated that they
cannot be individually resolved due to calorimeter detector segmentation and tracking failures.  

\begin{figure}[tb]
\begin{center}

\subfigure[$p_T(jet)>0.8$ TeV]{
 \includegraphics[scale=0.32, angle=0]{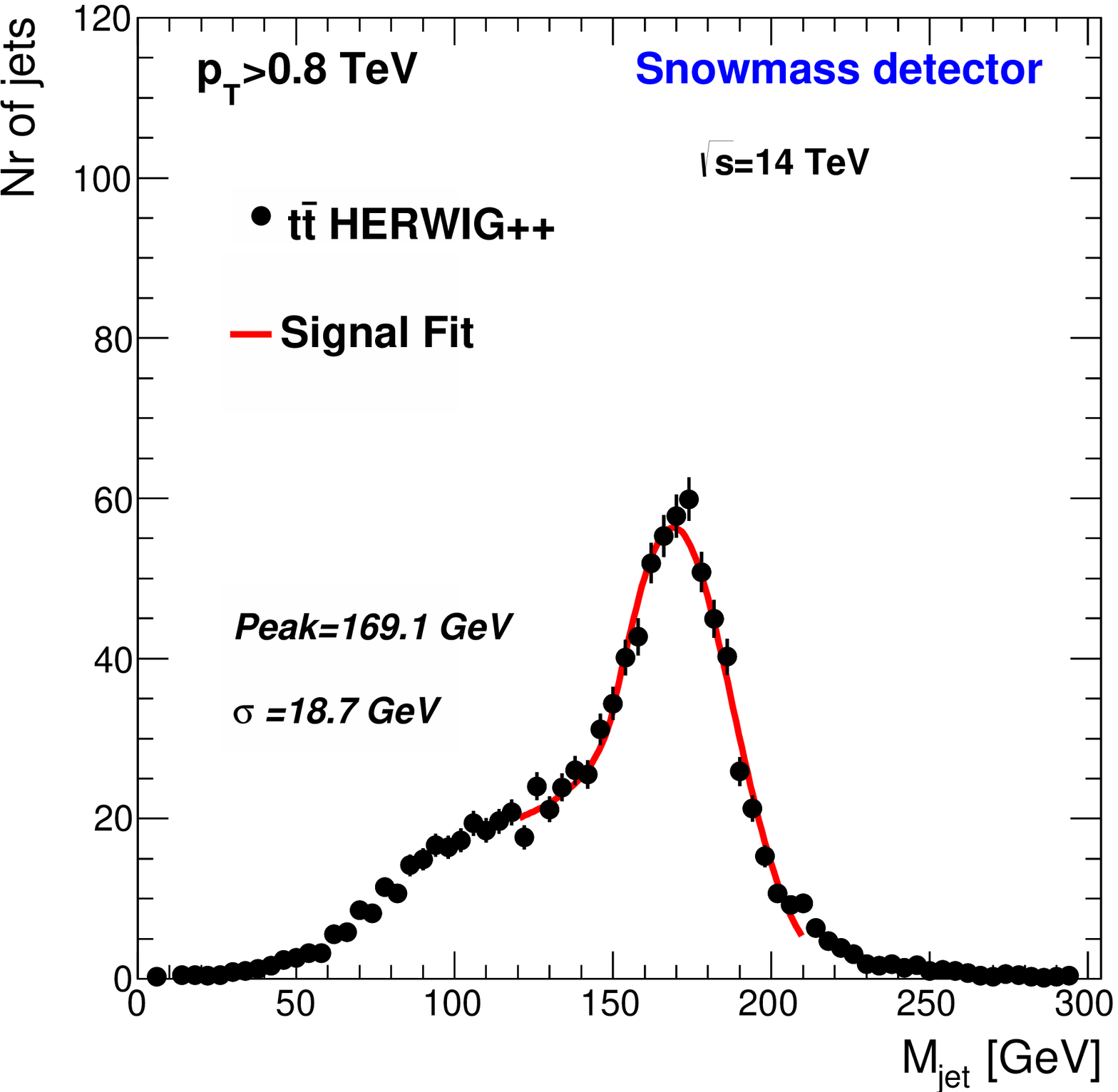}
 }
 \subfigure[$p_T(jet)>1.6$ TeV]{
 \includegraphics[scale=0.32, angle=0]{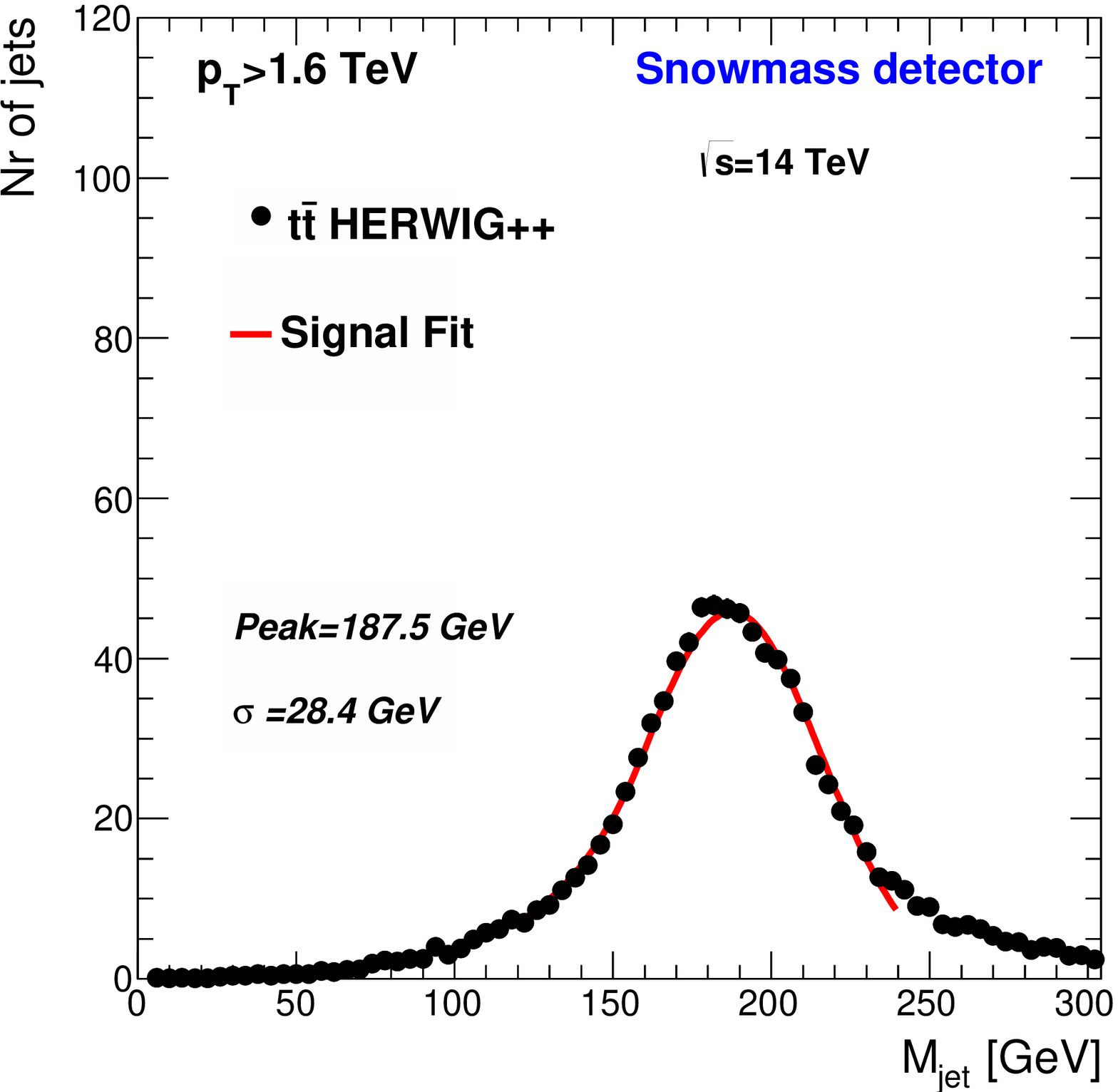}
 }
\end{center}
\caption{
Jet mass for $t\bar{t}$ events for different $p_{T}(\mathrm{jet})$ and $\langle \mu \rangle=140$.
The core of the peak was  fitted using a
Crystal Ball function~\cite{Oreglia}.
All histograms are normalized to 1000 events.
}
\label{fig:jetmass_pileupsFit}
\end{figure}

The effects mentioned above cause a degradation in the top quark jet resolution at large $p_T$.  For example, the width of the top quark jet mass peak increases by a factor of two when
comparing top quarks with $p_T > 1.6$ TeV to those with $p_T > 0.8$ TeV, see
Fig.~\ref{fig:jetmass_pileupsFit}. Algorithmic improvements extend the $p_T$ range where
top jets can be reconstructed, but ultimately the granularity of individual calorimeter cells
must be increased to maintain a good top jet reconstruction.

The reconstruction of top jets and substructure within large cone-size jets is a relatively
new field that has made tremendous progress in only a few years. More improvements are likely
to come, especially as sizable top quark event samples at the highest momenta become available
at the LHC. The ultimate limit is expected to come from the detector resolution, particularly
from calorimeter granularity, and future
detectors such as for CLIC or VLHC machines will need account for this.

\subsection{Lepton colliders}
~
A lepton collider (linear $e^+e^-$ colliders ILC and CLIC and circular $e^+e^-$ collider TLEP
and the $\mu^+\mu^-$ collider)  will allow for the study of electroweak production of
$t\overline{t}$ pairs with no concurring QCD background. Linear colliders can use polarized
beams, giving samples enriched in top quarks of left- or right-handed helicities. This can
allow one to probe new physics scenarios predicting anomalous production rates of 
right-handed $t$~quarks compared to the SM, and to disentangle the $t\gamma$ and $tZ$
couplings, see Section~\ref{sec:topcouplings}. 

Due to the electroweak production mechanism, all interesting processes occur at roughly the
same rate, and backgrounds can easily be reduced to a negligible level. After applying selection
cuts, it is possible to retain a signal sample of approximately 10$^5$ events at the 
500~GeV linear collider with 500~fb$^{-1}$ of integrated luminosity.
Unlike at the LHC, there are no or few additional interactions (pileup) per beam crossing,
especially for the ILC. Additional activity may come from $\gamma \gamma$ interactions.
Ongoing studies show that this residual pile-up can be controlled when applying the invariant
$k_t$ jet algorithm~\cite{Catani:1993hr,Ellis:1993tq} for background
suppression~\cite{WEUSTE:1499132}.

The lepton collider detectors can be designed to be more fine-grained and to have better resolution than the 
LHC detectors. The charge of the $b$~quark will be measured at a purity of 60\% and
better~\cite{Devetak:2010na}. This is indispensable for the measurement of $A^t_{FB}$ in fully
hadronic decays, see Section~\ref{sec:kinematics}. The jet energy resolution for LHC detectors
is between 10\% and 15\% for jets below 100~GeV~\cite{Aad:2012ag} whereas it is below 4\%
at the linear collider~\cite{Baer:2013cma}. This results in a clean top quark sample with
a narrow reconstructed mass as shown in Fig.~\ref{fig:TopMass}.

Using $A^t_{FB}$, the top-Higgs coupling $\lambda_t$ and the $t\overline{t}$ production 
cross-section, electroweak couplings can be determined at the percent level. It is important that
experimental and theoretical errors are kept at the same level. This requires a precise 
measurement of the luminosity and the beam polarization. Currently, both
parameters are expected to be controlled to better than 0.5\% at the linear collider.
In general the realization of machine and detectors must not compromise the precision physics.
This may be the biggest challenge in the coming years.

\section{Conclusions} 
\label{sec:topconclusions}

This is the concluding section for top quark snowmass 2013 studies. We have discussed six topics -- 
the top quark mass, top quark couplings to other SM particles, kinematics of top-like final states, 
rare decays of top quarks and top quark physics beyond the Standard Model. We will describe 
our conclusions for each of these topics. 

We have argued that a theoretically clean measurement of the top mass to about 300~MeV
is sufficient for many of the physics goals that are currently discussed, in particular
electroweak precision fits. If no new physics is
found at the LHC, it will be important to address the vacuum stability issue of the SM. To
address this, a top mass measurement with a precision of 100~MeV is required, given the
expected precision of the Higgs mass measurement.
The top quark mass can be measured with an accuracy of about 500~MeV in individual measurements
at the LHC, and their combination might reduce the uncertainty further. We note that both novel
methods and the high-luminosity option are required for achieving this accuracy.
The top mass can be measured with an accuracy of about 100~MeV
(dominated by theoretical uncertainties) at a lepton collider,
which matches well with the precision on the $W$~mass achievable at such a facility.

While the LHC and a future linear collider provide complementary information on top quark
couplings, there is no doubt that the LHC, especially the high-luminosity option,
will probe a majority of top quark
couplings to gluons, photons, $Z$'s, $W$'s and the Higgs boson with precision that should
allow us to detect deviations caused by generic BSM physics at the TeV scale.
The much higher precision achievable at a linear collider should then either allow us to study
these deviations or exclude the existence of generic BSM physics at even higher scales, in
particular for the $\gamma$ and $Z$~couplings.
The top Yukawa coupling, one of the most
important top couplings, will be measured to roughly equal precision at the LHC and the 500~GeV
ILC and to better precision at a high-energy linear collider.

Understanding how top quarks are produced and decay is an integral part of top physics at any collider.
Kinematic distributions and differential cross sections are the key to achieving this goal.
The measurement of basic top observables will help improve modeling of top quark events.
The large top event samples available in the future will allow the study of new observables
such as angular correlations or asymmetries that can uncover subtle new physics effects which
may not be accessible otherwise. We expect the LHC may be able to resolve the Tevatron $A_{FB}$ discrepancy.

The LHC and a future linear collider are complementary in probing rare decays of the top quark.
The LHC is better at probing flavor-changing couplings involving gluons, with about a factor two
improvement in the branching ratio limits expected from the high-luminosity option. A linear
collider is better for processes involving $\gamma$'s and $Z$'s. If rare decays are found, a
linear collider also is able to probe the spin structure of the couplings involved.

Top quarks play a very important role in searches for physics beyond the SM. In particular, 
solutions to the hierarchy problem require new particles decaying to top-like final states,
such as stops in SUSY or top partners in other models.
The LHC is able to cover the region of interest up to a few TeV in mass for stops, top-partners and
resonances decaying into top quarks. The high-luminosity option extends the mass reach for these
particles by roughly 50\%.
Given the current limits, only a multi-TeV lepton collider will be able to produce top partners and
resonances directly.
We note that there are stop models that might be difficult to discover at the LHC but can be probed
at a linear collider, for example stealth stops.

The 14~TeV LHC is a complex environment, especially the high pileup of the high-luminosity
option which makes precision measurements of top mass, couplings and kinematic
distributions challenging. Moreover, the 14~TeV LHC provides a large sample of boosted top quarks
for the first time whose decay products can no longer be individually identified using traditional techniques.
Our studies indicate that both of these challenges can be mitigated with algorithm developments
and other improvements, many of which have not been deployed yet for these Snowmass studies in the
high-luminosity scenario. 
In particular, many analyses will need to rely on these algorithms in future data collection periods, to maintain sensitivity to new physics processes in the high-mass regime
The experimental environment at a lepton collider does not suffer from
these problems and instead offers an ideal environment for precision top physics;
there are few or no additional interactions per crossing and the detectors are
more fine-grained and have better resolution.

In summary, the LHC and the HL-LHC will in two stages dramatically improve our knowledge of the top quark and extends the
reach for new physics to interesting and relevant regions. A future lepton collider will be able
to study the top quark in even more detail, in particular its mass and couplings.
We are confident that the predictions in this report are conservative and that the experiments
will do better with actual data than predicted here.


\bibliography{TopQuark/topquark}


\end{document}